\documentclass[12pt]{article}

\usepackage[utf8]{inputenc} 
\usepackage[T1]{fontenc}    

\usepackage{lmodern}

\usepackage{natbib}
\usepackage[english]{babel}

\usepackage{graphicx,amssymb,subfigure,icomma,xcolor,authblk}

\usepackage[labelfont=bf, width=\linewidth]{caption}

\usepackage{mathtools}
\usepackage[margin=1in]{geometry}

\usepackage{amsthm}
\usepackage{hyperref} 
\hypersetup{colorlinks=true, citecolor=blue, linkcolor=black, urlcolor =cyan}
\usepackage{hypcap}   
\usepackage{bookmark} 
\newtheorem{cor}{\corollaryname}[section]

\newtheorem{prop}[cor]{Proposition}

\numberwithin{equation}{section}

\newcommand{\minh}[1]{\colorbox{violet}{\color{white}   \textsf{\textbf{Minh}}} \textcolor{violet}{#1}}
\usepackage{multicol, booktabs, float, derivative, subcaption}
\newtheorem{lemma}{Lemma}[section]
\usepackage{array}
\newcolumntype{H}{>{\setbox0=\hbox\bgroup}c<{\egroup}@{}}

\newcommand{\DemandIndexSet}{I}
\newcommand{\DemandIndex}{i}
\newcommand{\FacilityIndexSet}{J}
\newcommand{\FacilityIndex}{j}
\newcommand{\FacilityAnotherIndex}{l}
\newcommand{\DemandVolume}{d}

\newcommand{\AvailableFacilityIndexSet}{J_1}
\newcommand{\CompetitorsFacilityIndexSet}{J'}
\newcommand{\AllFacilityIndexSet}{\bar{J}}

\newcommand{\LinearisationIndex}{n}
\newcommand{\LinearisationNIndices}{N}

\newcommand{\ConstraintIndex}{l}
\newcommand{\NConstraints}{m}

\newcommand{\UserEquilibrium}{\mathcal{Y}}

\newcommand{\Build}{x}
\newcommand{\BuildSet}{\mathcal{X}}
\newcommand{\DisaggregatedArrivalRate}{y}

\newcommand{\ArrivalRate}{\lambda}
\newcommand{\ServiceRate}{\mu}
\newcommand{\FacilityUtility}{U}
\newcommand{\FacilityRepresentativeTaste}{v}
\newcommand{\TravelTime}{t}
\newcommand{\WaitingTime}{\bar{w}}
\newcommand{\BalkingProbability}{\bar{p}}
\newcommand{\EffectiveArrivalRateFunction}{\bar{f}}
\newcommand{\EffectiveArrivalRate}{\bar{\lambda}}

\newcommand{\LinearisedFunctionValue}{\phi}

\newcommand{\LinearisationPointAssignment}{\hat{\DisaggregatedArrivalRate}_{n}}
\newcommand{\LinearisationPointAggregatedAssignment}{\hat{\ArrivalRate}_{n}}
\newcommand{\LinearisationDummyVar}{z}

\newcommand{\NSHint}[1]{}

\newcommand{\ProbabilityOfNUsers}[1]{p^{(#1)}}
\newcommand{\NServers}{s}
\newcommand{\QueueingSystemSize}{K}
\newcommand{\QueueingSystemSizeWithAccent}{K'}
\newcommand{\NUsers}{n}
\newcommand{\AverageNUsers}{L}

\newcommand{\InductionMainK}{K}
\newcommand{\InductionDummyK}{K}
\usepackage{makecell}

\title{Competitive EV charging station location with queues}

\author[1]{The Minh Nguyen}
\author[1]{Nagisa Sugishita}
\author[1]{Margarida Carvalho}
\author[2]{Amira Dems}

\affil[1]{\small CIRRELT and Département d’informatique et de recherche opérationnelle, Université de Montréal, Montréal, H3T 1J4, Canada}
\affil[2]{\small Institut de Recherche d’Hydro-Québec, Varennes, Quebec J3X 1S1, Canada}

\date{ }

\begin{document}
\maketitle


\begin{abstract}
Electric vehicle (EV) public charging infrastructure planning faces significant challenges in competitive markets, where multiple service providers affect congestion and user behavior. This work extends existing modeling frameworks by incorporating the presence of competitors' stations and more realistic queueing systems.

First, we analyze three finite queueing systems, $M/M/1/K$, $M/M/s/K$, and $M/E_r/s/K$, with varying numbers of servers (charging outlets) and service time distributions, deriving analytic expressions for user behavior metrics. Second, we embed the queueing-based user behavior model into a bilevel program, where the upper level locates new charging stations to maximize accessibility (throughput), and the lower level captures users' station choices via a user equilibrium. Third, we apply a reformulation from competitive congested user-choice facility location models to approximately solve the bilevel problem and introduce a surrogate-based heuristic to enhance scalability. Fourth, we showcase our methodology on a real-world case study of an urban area in Montreal (Canada), offering managerial insights into how user-choice behavior assumptions and competition affect throughput and location decisions. The results demonstrate that our model yields (re)location strategies that outperform the existing network. More broadly, this approach provides a tool for incorporating charging service quality—through queueing metrics—and existing competition into station planning.
\end{abstract}

\vspace{0.3cm}
\noindent
\textit{Keywords: Bilevel optimization, User equilibrium, Queueing theory, Facility location, Electric vehicle} 

\section{Introduction}

As electric vehicle (EV) adoption accelerates globally—driven by climate policies and financial incentives—the availability of reliable and accessible public charging infrastructure has become a critical concern for urban planning~\citep{iea_global_ev_outlook_2024}. This is especially true in dense residential areas, where many households lack access to private garages or driveways and must rely on public charging. Without strategic deployment of charging stations, cities risk creating service gaps that could discourage further adoption and undermine decarbonization targets. In response, many governments, such as those of Quebec~\citep{quebecEVstrategy2023} and British Columbia~\citep{cleanbcPublicCharger2025}, have prioritized infrastructure expansion to support the shift toward EVs, recognizing that widespread and equitable access to charging is essential not only for meeting current demand but also for sustaining long-term behavioral change in transportation choices.

Locating public EV charging stations in residential urban areas requires careful attention to multiple interrelated factors. First, spatial patterns of demand must be identified, reflecting where users live, travel, and park. Second, station capacity—including the number of charging points and the expected service time—must be sufficient to handle peak usage and avoid excessive queuing. Third, accessibility plays a key role, encompassing both physical proximity and ease of use, such as parking availability or walkability. Finally, planners must account for existing infrastructure, especially the presence of competing stations operated by other providers. Overlooking this competitive landscape can lead to redundant deployments, service inefficiencies, and reduced utilization. Together, these elements shape how well a charging network serves urban EV users and supports the broader goals of sustainable mobility.

Prior research on the location of EV charging stations has primarily focused on objectives such as maximizing coverage~\citep[e.g.,][]{arslan2016benders, YANG2018189, sugishita2025fair}, minimizing installation or user travel costs~\citep[e.g.,][]{xie2018long,filippi2023incorporating,Kinay2023ChargingSL}, and promoting EV adoption~\citep[e.g.,][]{ANJOS2020263, Lamontagne2022OptimisingEV}. These approaches, many of which are framed as variants of the Facility Location Problem (FLP), offer valuable tools for large-scale planning. However, most existing models assume simplistic user behavior and overlook the fact that EV users are self-interested and make decisions depending on individual utilities. In particular, while some models incorporate congestion effects such as waiting times or delays, they typically do not consider the lack of available parking space to queue—an important factor that can significantly impact system accessibility and user satisfaction \citep{YANG2018189,Kinay2023ChargingSL}. Furthermore, the presence of competing service providers is often neglected, despite the reality that multiple independent operators frequently coexist in urban environments. Although a few recent studies have incorporated aspects of competition between service providers \citep{CALVOJURADO2024105719, guillet2025coordinating}, they generally do not account for congestion effects, particularly those modeled using queueing theory. As a result, these models present an overly optimistic and potentially myopic view when applied to the complex realities of urban EV infrastructure planning.

In this work, we extend congested and competitive facility location models—originally introduced by \cite{Marianov2008FacilityLF} and later advanced by \cite{dan19}—to explicitly capture the behavioral and operational complexities of EV charging in urban environments. Specifically, we incorporate user choice dynamics and market competition into the planning framework, recognizing that EV users make self-interested decisions based on perceived utility rather than simple proximity. More concretely, we formulate the problem as a bilevel optimization model, where the upper level (leader) is responsible for strategic location decisions, determining which charging stations to open. We do not model queue capacity (i.e., number of charging outlets or available waiting space) nor service rate (i.e., charging power), as these are often constrained by external factors such as space availability and local grid capacity. Moreover, allowing these parameters to vary would introduce significant modeling and computational complexity. In our case study, most existing stations are equipped with two charging outlets, making this a reasonable simplification. At the lower level (follower), a user equilibrium model captures the behavior of EV users, who select charging stations—either those established by the leader or by competitors—so as to maximize their individual utilities. To realistically model the users' utilities, we integrate performance metrics that we derive from queueing systems, namely average waiting times and balking probabilities. These metrics allow us to represent service unavailability due to limited capacity and queueing constraints. We further compare various queueing models to determine the most appropriate for our context, based on a trade-off between analytical complexity and representational accuracy. Additionally, our upper-level objective is driven by location decisions maximizing accessibility, reflecting not only physical distance but also effective service availability. To solve our nonlinear mixed-integer bilevel optimization problem, we apply piecewise-linear approximations to the lower level and then reformulate it as a single-level mixed-integer linear program by leveraging the optimality conditions of the approximated lower-level problem. We also describe a surrogate-based heuristic. In a real-world case study, we demonstrate that our methodology can improve public charging station accessibility with respect to the existing infrastructure, particularly under high congestion scenarios where increased buffer capacity amplifies its advantages. By combining user behavior, competition, and congestion into a unified planning framework, our approach aims to offer a realistic and practical tool for tactical infrastructure deployment in dense, multi-operator urban settings. 

The paper is organized as follows. In Section~\ref{sec:related-work}, we provide a literature review on the congested and competitive FLP as well as on demand modeling. It also surveys studies on EV charging station placement, encompassing both FLP-based models and alternative approaches. Section~\ref{sec:problem} presents the problem, and Section~\ref{sec:user-equil} introduces the user choice model along with different queueing systems. In Section~\ref{sec:bilevel_formulation_of_facility_location_problem}, we discuss the linearization model of our bilevel formulation and a heuristic model based on the lower-level problem. In Section~\ref{sec:casestudy}, we conduct a case study on the lower-level and upper-level problems for \textit{Le Plateau-Mont-Royal}, a residential area in Montreal (Canada). In Section~\ref{sec:conclusion}, we derive conclusions from our work and we describe future research directions opened by the work presented in this paper.

\section{Related work}\label{sec:related-work}

In this section, we review the literature related to our work, focusing on modeling the FLP under congestion and competition, particularly in the context of EV charging station planning. We begin with an overview of the classical FLP and then discuss two important extensions that incorporate capacity limitations: the congested FLP and the capacitated FLP.
The congested FLP employs queueing models to capture demand-dependent service delays, while the capacitated FLP imposes strict limits on the demand that can be served. We then examine the role of competition in FLPs, followed by a review of EV charging station planning models, which we categorize into queue-based and non-queue-based approaches. Finally, we identify the research gap and position our contributions within this body of work.

FLPs provide a natural framework for optimizing the placement of service infrastructure such as EV charging stations. Classical FLPs aim to minimize cost or maximize service coverage through joint decisions on facility placement and customer allocation \citep{Laporte2019}. However, in many modern applications, particularly in transportation, demand is not centrally assigned but instead driven by user choice. This has led to the development of user-choice-based FLPs, where customer assignments are governed by utility-maximizing behavior, typically modeled using discrete choice frameworks such as the multinomial logit (MNL) model \citep{mcfadden_conditional_1974}.

A critical extension of the classical FLP is the congested FLP, where user utility depends on endogenous factors such as queueing delays. Early contributions by \cite{Marianov2008FacilityLF} and later by \cite{dan19} incorporated queueing delays and balking probabilities into MNL-based models, enabling a more realistic representation of congestion. Our work builds on this foundation by adopting more flexible queueing systems, including multiple servers and Erlang-distributed service times instead of the commonly assumed exponential distribution—choices that better reflect the operational characteristics of urban EV charging stations \citep{aldahabreh}. In contrast, studies such as \cite{ABOOLIAN2025107004} and \cite{JALILIMARAND2024442} assume infinite queue capacity and therefore do not model balking behavior.

Capacitated FLPs also address facility service limits but impose hard constraints rather than soft, congestion-based penalties. While queueing-based congestion models allow demand overflow with degraded service, capacitated models prohibit it outright. The capacitated FLP in conjunction with the user choice model has been considered by~\cite{haase2013management} and~\cite{ulloa2024logistics}. Studies such as \cite{fischetti2016benders, fischetti2017redesigning} propose decomposition techniques for solving such problems, although scalability remains a challenge due to the loss of separability in the presence of hard capacity limits.

In the context of competitive FLPs, the literature distinguishes between static, sequential, and dynamic models \citep{PLASTRIA2001461}. Our work follows the static competition framework, where a central planner decides where to place new facilities within a network where both the planner and a competitor already operate some existing facilities. After the new facilities are added, users respond by choosing which facilities to patronize. This is appropriate for EV charging markets, where operators must anticipate user behavior after taking their actions. Related studies in static competition include \cite{Ma2020-iq} and \cite{LIN2023106175}, though they differ in their treatment of user choice and congestion. While these models offer valuable insights, their layered choice structures and added complexity may hinder practical implementation in real-world EV charging networks.

In line with our queue-based modeling approach, we categorize the EV charging station planning literature into queue-based and non-queue-based FLPs. Queue-based models can incorporate finite-capacity queueing systems, but often neglect balking behavior \citep[e.g.,][]{XIAO2020101317}—a distinctive feature of systems with limited waiting space—or rely on infinite-capacity queues \citep[e.g.,][]{Kinay2023ChargingSL}, which is unrealistic in the context of EV charging stations. Moreover, both \cite{XIAO2020101317} and \cite{Kinay2023ChargingSL} focus on minimizing infrastructure costs but do not incorporate a user equilibrium model to represent demand, limiting their ability to capture self-interested user behavior. Others, like \citet{YANG2018189}, use general service time distributions but omit queue capacity and competitive dynamics. Additional studies that assume infinite queue capacity and ignore competition include \cite{Zhang04032023} and \cite{Liu01032023}. The former uses a multi-objective framework to minimize both total cost and service tardiness—whereas cost minimization is not central to our work—and proposes a genetic algorithm to determine the location of stations. The latter assumes deterministic charging times and employs a greedy approach to increase the capacity
of existing stations rather than opening new ones. In contrast, we use mixed-integer linear
programming (MILP) to approximately solve our station location optimization problem, as
well as a surrogate-based heuristic.

Non-queue-based intracity models in the literature often present key limitations: some do not impose capacity constraints at all \citep[e.g.,][]{Lamontagne2022OptimisingEV, lamontagne2024accelerated}, others model capacity using simplified flow-based approaches \citep[e.g.,][]{Parent2023MaximumFF} that ignore congestion effects such as waiting and service times, and some assume centrally assigned demand rather than modeling self-interested user behavior \citep[e.g.,][]{filippi2023incorporating}. In intercity contexts, \cite{sugishita2025fair} model congestion using the well-known \emph{Bureau of Public Roads} (BPR) function instead of queueing theory. \cite{ANJOS2020263} is the only study to address both intracity and intercity planning simultaneously. Their approach combines node- and flow-based demand modeling with a rolling-horizon heuristic to solve a large-scale, multi-period problem under capacity constraints, defined as limits on how many users can be assigned to each station per time period. On the competition side, \cite{CALVOJURADO2024105719} develop siting strategies to avoid competitors, but do not account for congestion. \cite{guillet2025coordinating} consider a congestion game framework to model strategic interactions between EV navigation platforms. However, this refers to competition over shared resources, not physical congestion such as waiting times or queue lengths at charging stations. Moreover, their work does not address the problem of optimizing the locations of charging stations; instead, it focuses on a navigation platform that recommends users to available stations within an existing infrastructure network.

In contrast with previous research in EV charging station placement, our work models \emph{(i)} self-interested user behavior under congestion, using realistic queueing models, \emph{(ii)} accounts for existent stations, notably those of competitors, and \emph{(iii)} considers a novel throughput-based objective to guide location decisions. This allows us to capture key operational constraints of urban EV charging systems and inform infrastructure planning under realistic usage patterns. We then demonstrate the practical applicability of our model through a real-world case study in Le Plateau-Mont-Royal, a densely-populated residential neighborhood in Montreal.

\section{Problem statement}\label{sec:problem}

We consider the decision-making process of EV users when selecting charging stations to recharge their vehicles. We model this as a bipartite graph, where demand nodes (e.g., user origins or population centers) are connected to facility nodes (charging stations). This setting captures two key features: congestion, which arises at the facility level when excessive demand exceeds service capacity (e.g., limited charging outlets or waiting space), and competition, as users may choose among stations operated by different service providers.

Let $\DemandIndexSet$ be the set of demands (representing populations grouped or aggregated by their geographical locations), $\FacilityIndexSet$ be the set of existing and potential facilities. For each facility $\FacilityIndex \in \FacilityIndexSet$, we use binary variable $\Build_{\FacilityIndex}$ to indicate the availability of facility $\FacilityIndex$, and let $\BuildSet \subset \{0, 1\}^{|\FacilityIndexSet|}$ be the set of feasible $\Build$. 
We can model existing facility $\FacilityIndex$ by fixing $\Build_{\FacilityIndex} = 1$ for all $\Build \in \BuildSet$.
We also define $\CompetitorsFacilityIndexSet$ as the set of existing competitors' facilities and $\AllFacilityIndexSet = \FacilityIndexSet \cup \CompetitorsFacilityIndexSet$ as the set of all facilities.
Given $\Build \in \BuildSet$, users choose which facilities to patronize based on individual utilities, which we define in the next section.
For each demand $\DemandIndex \in \DemandIndexSet$ and facilities $\FacilityIndex \in \AllFacilityIndexSet$, let $\DisaggregatedArrivalRate_{\DemandIndex \FacilityIndex}$ be the share of demand $\DemandIndex$ that patronizes facility $\FacilityIndex$, and $\DemandVolume_{\DemandIndex}$ be the demand volume of $\DemandIndex$.

The facility location problem that maximizes the throughput of the operator's facilities can be written as
\begin{align}
\label{eq:abstract_flp}
  \max_{\Build, \DisaggregatedArrivalRate} \ & \EffectiveArrivalRateFunction\left( 
  \DisaggregatedArrivalRate \right)
  \\
  \text{s.t.}\ & \Build \in \BuildSet, \DisaggregatedArrivalRate \in \UserEquilibrium(\Build), \notag
\end{align}
where $\EffectiveArrivalRateFunction$ is a function that maps user choices $\DisaggregatedArrivalRate$ to the throughput of the facilities (a more concrete form will be described in Section~\ref{sec:bilevel_formulation_of_facility_location_problem}) and $\UserEquilibrium(\Build)$ is the set of the user assignment given $\Build$.
In this work, we assume $\DisaggregatedArrivalRate$ is given as a user equilibrium, as discussed in the next section.

\section{User equilibrium}\label{sec:user-equil}

In this section, we describe our demand model at available facilities, i.e., the set $\UserEquilibrium(\Build)$ for all $\Build \in \BuildSet$. To this end, in Section~\ref{subsec:userchoice}, we review the usage of user choice models to guide the selection of facilities based on users' utilities. 
Since our (dis)utility functions incorporate the metrics of waiting time and balking probability at facilities, in Section~\ref{subsec:queueingtheory}, we derive these metrics using queueing models of practical interest. We analyze two relevant systems, denoted as $M/M/s/K$ and $M/E_r/s/K$, where the first term refers to the arrival process ($M$ for memoryless, i.e., Poisson arrivals), the second to the service time distribution ($M$ for exponential, $E_r$ for Erlang of shape $r$), $s$ is the number of servers (charging outlets), and $K$ is the system capacity, including both servers and buffer (i.e., parking spots). While the derivation itself follows standard theory, we show that the $M/M/s/K$ model closely approximates the more general simulated $M/E_r/s/K$ model  in our setting, which is a novel insight of our analysis. In Section~\ref{subsec:mmsk}, we present our main methodological contribution at the lower level: integrating the metrics of $M/M/s/K$ into the lower-level (user equilibrium) problem and approximate it through a linear program.

\subsection{User choice model}\label{subsec:userchoice}

In this section, we outline our user choice model that describes the user behaviors: Given a set of available facilities $\AvailableFacilityIndexSet$ defined as $\{ \FacilityIndex \in \FacilityIndexSet : \Build_{\FacilityIndex} = 1 \}$ and a set of competitors facilities $\CompetitorsFacilityIndexSet$, which facilities are chosen by users?
We assume that each user's choice probabilities over the facilities follow the MNL model by~\citet{mcfadden_conditional_1974}. The MNL model allows for a more realistic representation of user behavior compared to the commonly used simplified model in which users are assumed to choose the closest stations~\citep{Kinay2023ChargingSL, Parent2023MaximumFF,Lamontagne2022OptimisingEV}. We outline this model below.

To simplify the exposition, for the moment, we assume that the expected waiting time and the probability of balking at facility $\FacilityIndex$ are fixed and known to be $\WaitingTime_{\FacilityIndex}$ and $\BalkingProbability_{\FacilityIndex}$, respectively, for each facility $\FacilityIndex \in \AvailableFacilityIndexSet \cup \CompetitorsFacilityIndexSet$. Although incorporating $\CompetitorsFacilityIndexSet$ into the user choice model introduces a strong assumption regarding the modeling of competition, this approach is consistent with previous literature. Furthermore, key information about competitor charging infrastructure—such as charging power and the number of outlets—is typically public and accessible to users. Even when full specifications are not disclosed, such details can often be inferred from observable station characteristics.
As is done by \citet{dan19}, we further assume that the disutility of facility $\FacilityIndex$ for a user from demand $\DemandIndex$ is a random variable given by
\begin{equation}
\label{eq:definition_ofutility}
\FacilityUtility_{\DemandIndex \FacilityIndex} = \FacilityRepresentativeTaste_{\DemandIndex \FacilityIndex} - \varepsilon_{\DemandIndex \FacilityIndex},
\end{equation}
where $\varepsilon_{\DemandIndex \FacilityIndex}$ is a random variable that follows the Gumbel distribution with common scale parameter $\theta$ and variance $\dfrac{\pi ^2}{6\theta^2}$, and
$$
\FacilityRepresentativeTaste_{\DemandIndex \FacilityIndex} = \TravelTime_{\DemandIndex \FacilityIndex} + \alpha \WaitingTime_{\FacilityIndex} + \beta \BalkingProbability_{\FacilityIndex} 
$$
with some nonnegative parameters $\alpha$ and $\beta$.
Under this setup, \citet{mcfadden_conditional_1974} shows that the probability of a user randomly sampled from demand $\DemandIndex$ to choose facility $\FacilityIndex \in \AvailableFacilityIndexSet \cup \CompetitorsFacilityIndexSet$ is given by
$$
\frac{
\displaystyle
 e^{-\theta \FacilityRepresentativeTaste_{\DemandIndex \FacilityIndex}}
}{
\displaystyle
\sum_{\FacilityAnotherIndex \in \AvailableFacilityIndexSet} e^{-\theta \FacilityRepresentativeTaste_{\DemandIndex \FacilityAnotherIndex}}
+
\sum_{\FacilityAnotherIndex \in \CompetitorsFacilityIndexSet} e^{-\theta \FacilityRepresentativeTaste_{\DemandIndex \FacilityAnotherIndex}}
}.
$$
Note that $\theta$ is a parameter that controls the stochasticity in users' behaviors.
Larger $\theta$ magnifies the impact of $\FacilityRepresentativeTaste_{\DemandIndex \FacilityIndex}$ on users' assignments.

Above, we assume the expected waiting time and the probability of balking to be fixed.
However, usually, these quantities depend on the amount of demand assigned to the facility (i.e., the amount of demand patronizing the facility).
Suppose that for some demand $\DemandIndex \in \DemandIndexSet$, there is an available facility $\FacilityIndex \in \AvailableFacilityIndexSet \cup \CompetitorsFacilityIndexSet$ nearby.
This facility $\FacilityIndex$ is attractive for demand $\DemandIndex$ in the sense that it has a small value of $\TravelTime_{\DemandIndex \FacilityIndex}$.
However, if the entire population of demand $\DemandIndex$ patronizes facility $\FacilityIndex$, it may cause significant congestion at the facility.
This results in large values of expected waiting time $\WaitingTime_{\FacilityIndex}$ and balking probability $\BalkingProbability_{\DemandIndex \FacilityIndex}$, in turn degrading the utility of facility $\FacilityIndex$ and motivating the users to choose other facilities.

To model this situation, \cite{Marianov2008FacilityLF} and \cite{dan19} consider the user assignment given as a solution to the following system of equations:
\begin{align}
\label{eq:user_equilibirum_fixed_point_formulation}
\DisaggregatedArrivalRate_{\DemandIndex \FacilityIndex} &= \DemandVolume_{\DemandIndex} \frac{
\displaystyle
e^{-\theta \FacilityRepresentativeTaste_{\DemandIndex \FacilityIndex}}
}{
\displaystyle
\sum_{\FacilityAnotherIndex \in \AvailableFacilityIndexSet} e^{-\theta \FacilityRepresentativeTaste_{\DemandIndex \FacilityAnotherIndex}}
+
\sum_{\FacilityAnotherIndex \in \CompetitorsFacilityIndexSet} e^{-\theta \FacilityRepresentativeTaste_{\DemandIndex \FacilityAnotherIndex}}
}, 
&& \qquad \forall \DemandIndex \in \DemandIndexSet, \FacilityIndex \in \AvailableFacilityIndexSet \cup \CompetitorsFacilityIndexSet,
\\
\FacilityRepresentativeTaste_{\DemandIndex \FacilityIndex} &= \TravelTime_{\DemandIndex \FacilityIndex} + \alpha \WaitingTime_{\FacilityIndex}(\ArrivalRate_{\FacilityIndex}) + \beta \BalkingProbability_{\FacilityIndex}(\ArrivalRate_{\FacilityIndex}), 
&& \qquad \forall \DemandIndex \in \DemandIndexSet, \FacilityIndex \in \AvailableFacilityIndexSet \cup \CompetitorsFacilityIndexSet,
\notag \\
\ArrivalRate_{\FacilityIndex} &= \sum_{\DemandIndex \in \DemandIndexSet} \DisaggregatedArrivalRate_{\DemandIndex \FacilityIndex}, 
&& \qquad \forall \FacilityIndex \in \AvailableFacilityIndexSet \cup \CompetitorsFacilityIndexSet,
\notag \\
\DisaggregatedArrivalRate_{\DemandIndex \FacilityIndex} &= 0, && \qquad \forall \DemandIndex \in \DemandIndexSet, \FacilityIndex \in \FacilityIndexSet \setminus \AvailableFacilityIndexSet, \notag
\end{align}
where $\DemandVolume_{\DemandIndex}$ is the volume of demand $\DemandIndex \in \DemandIndexSet$ and $\WaitingTime(\ArrivalRate_{\FacilityIndex})$/$\BalkingProbability(\ArrivalRate_{\DemandIndex})$ is the expected waiting time/balking probability at facility $\FacilityIndex$ given user assignment $\ArrivalRate_{\FacilityIndex}$ at facility $\FacilityIndex$.
\cite{Marianov2008FacilityLF} show the existence of the solution to this system of equations under very mild assumptions, which are satisfied in our model.

\cite{fisk80} shows that the set of solutions to~\eqref{eq:user_equilibirum_fixed_point_formulation} coincides with the optimal solution set for the following optimization problem:
\begin{alignat}{2}
\label{eq:llp}
\tag{LLP($\Build$)}
\min_{\DisaggregatedArrivalRate, \ArrivalRate} \ & \mathmakebox[0.7\textwidth][l]{\sum_{\DemandIndex \in \DemandIndexSet} \sum_{\FacilityIndex \in \AllFacilityIndexSet} \left(
\dfrac{1}{\theta} \DisaggregatedArrivalRate_{\DemandIndex \FacilityIndex} \ln \DisaggregatedArrivalRate_{\DemandIndex \FacilityIndex} + \DisaggregatedArrivalRate_{\DemandIndex \FacilityIndex} \TravelTime_{\DemandIndex \FacilityIndex} 
\right) 
+ \sum_{\FacilityIndex \in \AllFacilityIndexSet} \left( \alpha \int_0^{\ArrivalRate_{\FacilityIndex}} \WaitingTime_{\FacilityIndex}(q)\,dq 
+ \beta \int_0^{\ArrivalRate_{\FacilityIndex}} \BalkingProbability_{\FacilityIndex}(q)\,dq \right)} \\
\text{s.t.} \ & 
\sum_{\FacilityIndex \in \AllFacilityIndexSet} \DisaggregatedArrivalRate_{\DemandIndex \FacilityIndex} = \DemandVolume_{\DemandIndex}, &&\forall \DemandIndex \in \DemandIndexSet \notag, \\
& \ArrivalRate_{\FacilityIndex} = \sum_{\DemandIndex \in \DemandIndexSet} \DisaggregatedArrivalRate_{\DemandIndex \FacilityIndex}, &&\forall \FacilityIndex \in \AllFacilityIndexSet, \notag \\
& \DisaggregatedArrivalRate_{\DemandIndex \FacilityIndex} \le \DemandVolume_{\DemandIndex} \Build_{\FacilityIndex}, && \forall \DemandIndex \in \DemandIndexSet, \FacilityIndex \in \FacilityIndexSet, \notag \\
& \DisaggregatedArrivalRate_{\DemandIndex \FacilityIndex} \geq 0,  &&\forall \DemandIndex \in \DemandIndexSet, \FacilityIndex \in \AllFacilityIndexSet. \notag
\end{alignat}

So far, we have assumed that the utility of a user is a random variable and discussed the MNL model. When the randomness in the utility~\eqref{eq:definition_ofutility} is set to zero, i.e., $\varepsilon_{\DemandIndex \FacilityIndex} = 0$, the resulting deterministic behavior corresponds to a Wardrop equilibrium~\citep{BeckmannEtAl1955}. It is of interest to observe that the set of the Wardrop equilibrium corresponds to the solution set of~\eqref{eq:llp} with $1 / \theta = 0$:
\begin{alignat*}{2}
\min_{\DisaggregatedArrivalRate, \ArrivalRate} \ & \mathmakebox[0.7\textwidth][l]{\sum_{\DemandIndex \in \DemandIndexSet} \sum_{\FacilityIndex \in \AllFacilityIndexSet}
\DisaggregatedArrivalRate_{\DemandIndex \FacilityIndex} \TravelTime_{\DemandIndex \FacilityIndex}
+ \sum_{\FacilityIndex \in \AllFacilityIndexSet} \left( \alpha \int_0^{\ArrivalRate_{\FacilityIndex}} \WaitingTime_{\FacilityIndex}(q)\,dq 
+ \beta \int_0^{\ArrivalRate_{\FacilityIndex}} \BalkingProbability_{\FacilityIndex}(q)\,dq \right)} \\
\text{s.t.} \ & 
\sum_{\FacilityIndex \in \AllFacilityIndexSet} \DisaggregatedArrivalRate_{\DemandIndex \FacilityIndex} = \DemandVolume_{\DemandIndex}, &&\forall \DemandIndex \in \DemandIndexSet \notag, \\
& \ArrivalRate_{\FacilityIndex} = \sum_{\DemandIndex \in \DemandIndexSet} \DisaggregatedArrivalRate_{\DemandIndex \FacilityIndex}, &&\forall \FacilityIndex \in \AllFacilityIndexSet, \notag \\
& \DisaggregatedArrivalRate_{\DemandIndex \FacilityIndex} \le \DemandVolume_{\DemandIndex} \Build_{\FacilityIndex}, && \forall \DemandIndex \in \DemandIndexSet, \FacilityIndex \in \FacilityIndexSet, \notag \\
& \DisaggregatedArrivalRate_{\DemandIndex \FacilityIndex} \geq 0,  &&\forall \DemandIndex \in \DemandIndexSet, \FacilityIndex \in \AllFacilityIndexSet. \notag
\end{alignat*}

Lastly, the waiting time $\WaitingTime_{\FacilityIndex}$ and the balking probability $\BalkingProbability_{\FacilityIndex}$ depend on the underlying queue.
In the next section, we review three types of queues, discuss how to obtain or approximate these quantities and compare their results.

\subsection{Queueing theory}\label{subsec:queueingtheory}

As stated in the previous section, we assume each facility is modeled as a queue.
Furthermore, the utility perceived by a user depends on the congestion level at the facility, in particular the waiting time and the balking probability.
The appropriate choice of the queue depends on the nature of the facilities being modeled.

\cite{dan19} assume that the underlying queues are $M/M/1/\QueueingSystemSize$ queues, a single-server queue with exponentially distributed service time.
In this work, we extend their work and consider $M/M/\NServers/\QueueingSystemSize$ queues, a multi-server queue with exponentially distributed service time.
Compared with the $M/M/1/\QueueingSystemSize$ queue, the $M/M/\NServers/\QueueingSystemSize$ queue offers great flexibility to model various types of facilities.
For example,~\cite{Smith2008MGCkPM} demonstrates that a $M/E_r/\NServers/\QueueingSystemSize$ queue, where service times follow the Erlang-$r$ distribution, can be approximated with a $M/M/\NServers/\QueueingSystemSizeWithAccent$ queue to a high accuracy with appropriately chosen $\QueueingSystemSizeWithAccent$.

In this section, we review the $M/M/\NServers/\QueueingSystemSize$ queue (Section~\ref{subsec:mmsk_queue}) and the method of~\cite{Smith2008MGCkPM} to approximate the $M/E_r/\NServers/\QueueingSystemSize$ queue with the $M/M/\NServers/\QueueingSystemSizeWithAccent$ queue (Section~\ref{subsec:mesk_queue}). In addition, we compare the performance metrics—balking probability and average waiting time—of these queues (Section~\ref{subsec:queues_comparison}).


\subsubsection{$M/M/\NServers/\QueueingSystemSize$ queue}\label{subsec:mmsk_queue}

Performance metric formulas for the $M/M/\NServers/\QueueingSystemSize$ queue are well-known and can be found in a standard textbook.
For example, see Section 11.6 of~\cite{stewart09}.
In this section, we briefly review the results relevant to our work.

Let us denote the arrival rate by $\ArrivalRate$, the service rate of each server by $\ServiceRate$, the number of servers by $\NServers$, and the size of the queueing system (the sum of the number of servers and the number of waiting spaces) by $\QueueingSystemSize$.
The probability $\ProbabilityOfNUsers{\NUsers}$ of having $\NUsers$ users in the system, the number of users who are being served and who are waiting, is
\begin{align}
\label{eq:mmsk_probability_of_n_users}
\ProbabilityOfNUsers{\NUsers} &= 
\begin{cases}
\dfrac{1}{Z \NUsers!} \left(\dfrac{\ArrivalRate}{\ServiceRate} \right)^{\NUsers},
&
\quad 0 \leq \NUsers \leq \NServers, \\
\dfrac{1}{Z \NServers^{\NUsers - \NServers} \NServers!}
\left(\dfrac{\ArrivalRate}{\ServiceRate} \right)^{\NUsers},
&
\quad \NServers \leq \NUsers \leq \QueueingSystemSize, \\
0, &\quad \NUsers > \QueueingSystemSize,
        \end{cases}
\end{align}
where $Z$ is the normalization constant given as
\begin{align*}
Z
&=
\sum_{m = 0}^{\NServers - 1} \dfrac{1}{m!} \left(\dfrac{\ArrivalRate}{\ServiceRate} \right)^{m}
+
\sum_{m = \NServers}^{\QueueingSystemSize} \dfrac{1}{\NServers^{m - \NServers} \NServers!}
\left(\dfrac{\ArrivalRate}{\ServiceRate} \right)^{m}
\\
&=
\begin{cases}
\displaystyle
\sum_{m = 0}^{\NServers - 1} \dfrac{1}{m!} \left(\dfrac{\ArrivalRate}{\ServiceRate} \right)^{m}
+
\dfrac{\NServers^\NServers (\QueueingSystemSize - \NServers + 1)}{\NServers!},
&
\text{ if $\ArrivalRate = \NServers \ServiceRate$,}
\\
\displaystyle
\sum_{m = 0}^{\NServers - 1} \dfrac{1}{m!} \left(\dfrac{\ArrivalRate}{\ServiceRate} \right)^{m}
+
\dfrac{
(\NServers \rho)^\NServers
 (1 - \rho^{\QueueingSystemSize - \NServers + 1})}{\NServers! (1 - \rho)},
&
\text{ otherwise,}
\end{cases}
\end{align*}
with $\rho = \ArrivalRate / (\NServers \ServiceRate)$.
In particular, we have the balking probability $\BalkingProbability = \ProbabilityOfNUsers{\QueueingSystemSize}$.

Using~\eqref{eq:mmsk_probability_of_n_users}, one can compute the average number of users in the system as
$$
\AverageNUsers = 
\sum_{m = 0}^\QueueingSystemSize m \, \ProbabilityOfNUsers{m}.
$$
It follows from Little's formula that the average waiting time is given by
\begin{align}
\label{eq:queueing_average_waiting_time}
\WaitingTime = \dfrac{\AverageNUsers}{\ArrivalRate (1 - \ProbabilityOfNUsers{\QueueingSystemSize})}
=
\dfrac{1}{\ArrivalRate (1 - \ProbabilityOfNUsers{\QueueingSystemSize})}
\sum_{m = 0}^\QueueingSystemSize m \, \ProbabilityOfNUsers{m}.
\end{align}

\subsubsection{$M/E_r/\NServers/\QueueingSystemSize$ queue}\label{subsec:mesk_queue}

The $M/E_r/\NServers/\QueueingSystemSize$ queue assumes that the service time follows the Erlang-$r$ distribution.
To the best of our knowledge, due to the complexity of multiple servers and Erlang-$r$ distributed service times, there is no exact method to find the performance measures (e.g., the balking probability and the expected waiting time) of a $M/E_r/\NServers/\QueueingSystemSize$ queue.

One popular approach is the so-called the two-moment method of~\cite{Smith2008MGCkPM} to approximate $M/E_r/\NServers/\QueueingSystemSize$ with $M/M/\NServers/\QueueingSystemSizeWithAccent$ where $\QueueingSystemSizeWithAccent$ is a constant given by
\begin{align*}
    \QueueingSystemSizeWithAccent = \dfrac{\QueueingSystemSize - \NServers}{1+T} + \NServers
\end{align*}
where
\begin{align*}
    T = \dfrac{1}{2} \left(\dfrac{1}{r} - 1\right) \sqrt{\dfrac{\ArrivalRate}{\NServers \ServiceRate} \exp\left(-\dfrac{1}{r}\right)}.
\end{align*}

Now, by substituting $\QueueingSystemSizeWithAccent$ for $\QueueingSystemSize$ in \eqref{eq:mmsk_probability_of_n_users}, we obtain an approximation of the probability $\ProbabilityOfNUsers{\NUsers}$ of having $\NUsers$ users in the system as
\begin{align}
\label{eq:mesk_probability_of_n_users}
\ProbabilityOfNUsers{\NUsers} &\approx
\begin{cases}
\dfrac{1}{Z' \NUsers!} \left(\dfrac{\ArrivalRate}{\ServiceRate} \right)^{\NUsers},
&
\quad 0 \leq \NUsers \leq \NServers, \\
\dfrac{1}{Z' \NServers^{\NUsers - \NServers} \NServers!}
\left(\dfrac{\ArrivalRate}{\ServiceRate} \right)^{\NUsers},
&
\quad \NServers \leq \NUsers \leq \QueueingSystemSize, \\
0, &\quad \NUsers > \QueueingSystemSize,
        \end{cases}
\end{align}
where $Z'$ is the ``normalization'' constant given as
\begin{align*}
Z'
&=
\begin{cases}
\displaystyle
\sum_{m = 0}^{\NServers - 1} \dfrac{1}{m!} \left(\dfrac{\ArrivalRate}{\ServiceRate} \right)^{m}
+
\dfrac{\NServers^\NServers (\QueueingSystemSizeWithAccent - \NServers + 1)}{\NServers!},
&
\text{ if $\ArrivalRate = \NServers \ServiceRate$,}
\\
\displaystyle
\sum_{m = 0}^{\NServers - 1} \dfrac{1}{m!} \left(\dfrac{\ArrivalRate}{\ServiceRate} \right)^{m}
+
\dfrac{
(\NServers \rho)^\NServers
 (1 - \rho^{\QueueingSystemSizeWithAccent - \NServers + 1})}{\NServers! (1 - \rho)},
&
\text{ otherwise,}
\end{cases}
\end{align*}
with $\rho = \ArrivalRate / (\NServers \ServiceRate)$.
The average waiting time can be approximated by using~\eqref{eq:mesk_probability_of_n_users} and~\eqref{eq:queueing_average_waiting_time}.

\subsubsection{Queues comparison} \label{subsec:queues_comparison}

In this section, we briefly compare the queues numerically.
The goal of this section is three-fold: to observe the difference between 1) the single-server $M/M/1/K$ queue and the multi-server $M/M/s/K$ queue with $s \ge 2$, 2) the $M/M/s/K$ queue and the $M/E_r/s/K$ queue, and 3) the $M/E_r/s/K$ queue and the $M/M/s/K'$ queue to approximate the $M/E_r/s/K$ queue using the method of \cite{Smith2008MGCkPM}.

To this end, we consider four types of queues.
Let $B$ be an integer indicating the buffer size, that is, the capacity of the queue $\QueueingSystemSize$ minus the number of servers $s$. 
The first queue is the $M/M/1/1+B$ queue with a single server of service rate 40.
The system capacity $1 + B$ is the number of servers plus the buffer size.
The second queue is the $M/M/2/2+B$ queue with two servers.
Each server's service rate is set to 20 so the total service rate of the system is 40.
The third queue is the $M/E_2/2/2+B$ queue with two servers.
The service time of each server follows the Erlang-2 distribution with a service rate of 20.
Lastly, the fourth queue is the $M/M/2/2+B'$ queue, which is obtained to approximate the $M/E_2/2/2+B$ queue with the method of~\cite{Smith2008MGCkPM}.
Once again, the servers have a service rate of 20 each.
The four queues are summarized in Table~\ref{tab:queues}


\begin{table}[hbtp]
\begin{tabular}{ccccc}
\toprule
Queue & Service time distribution & \# of servers & Total service rate & Buffer size \\
\midrule
$M/M/1/1+B$ & Exponential & 1 & 40 & $B$ \\
$M/M/2/2+B$ & Exponential & 2 & 40 & $B$ \\
$M/E_2/2/2+B$ & Erlang-2 & 2 & 40 & $B$ \\
$M/M/2/2+B'$ & Exponential & 2 & 40 & $B'$ \\
\bottomrule
\end{tabular}
\caption{The four types of queues compared in this section}
\label{tab:queues}
\end{table}

\begin{figure}[h]
    \centering
     \includegraphics[width=\linewidth]{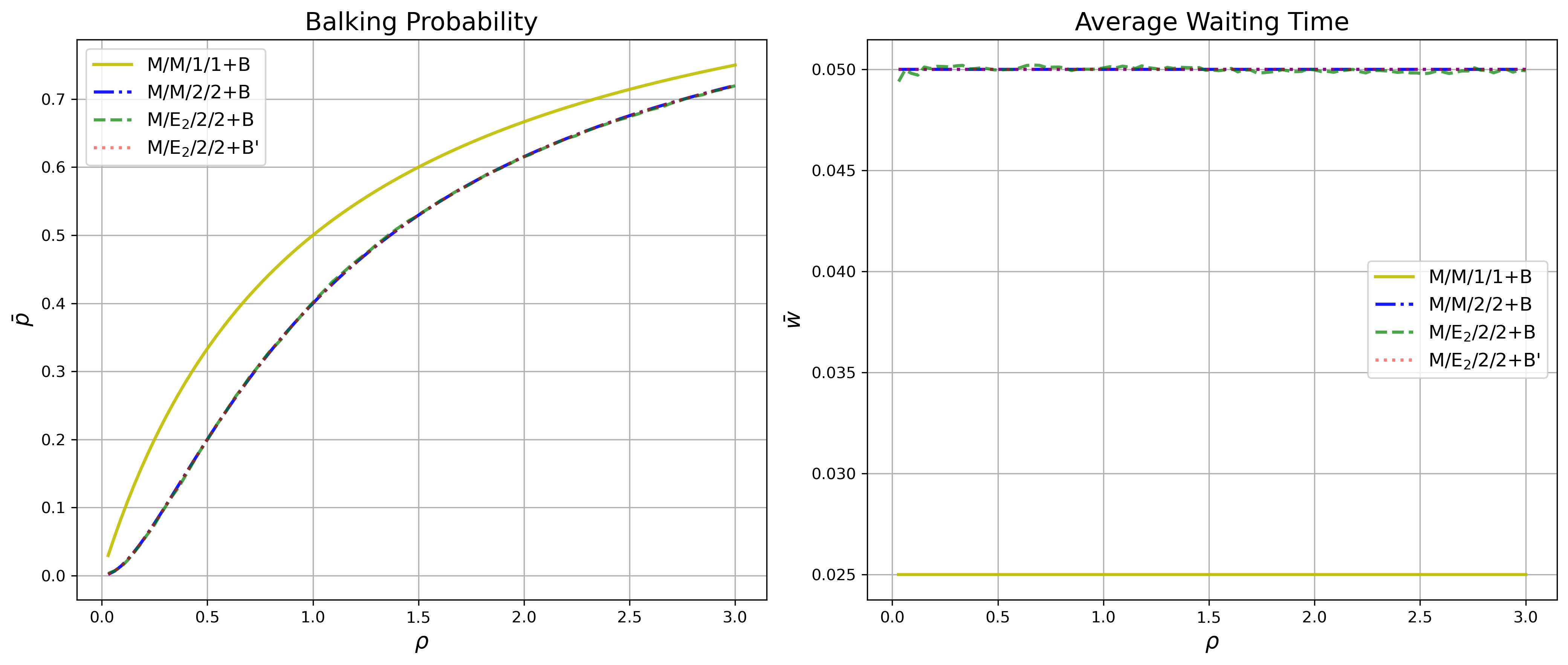}
    \caption{Queue performance comparison for $B=0$.
    }\label{fig:comparison_b0_s=2}
\end{figure}
\begin{figure}[h]
    \centering
    \includegraphics[width=\linewidth]{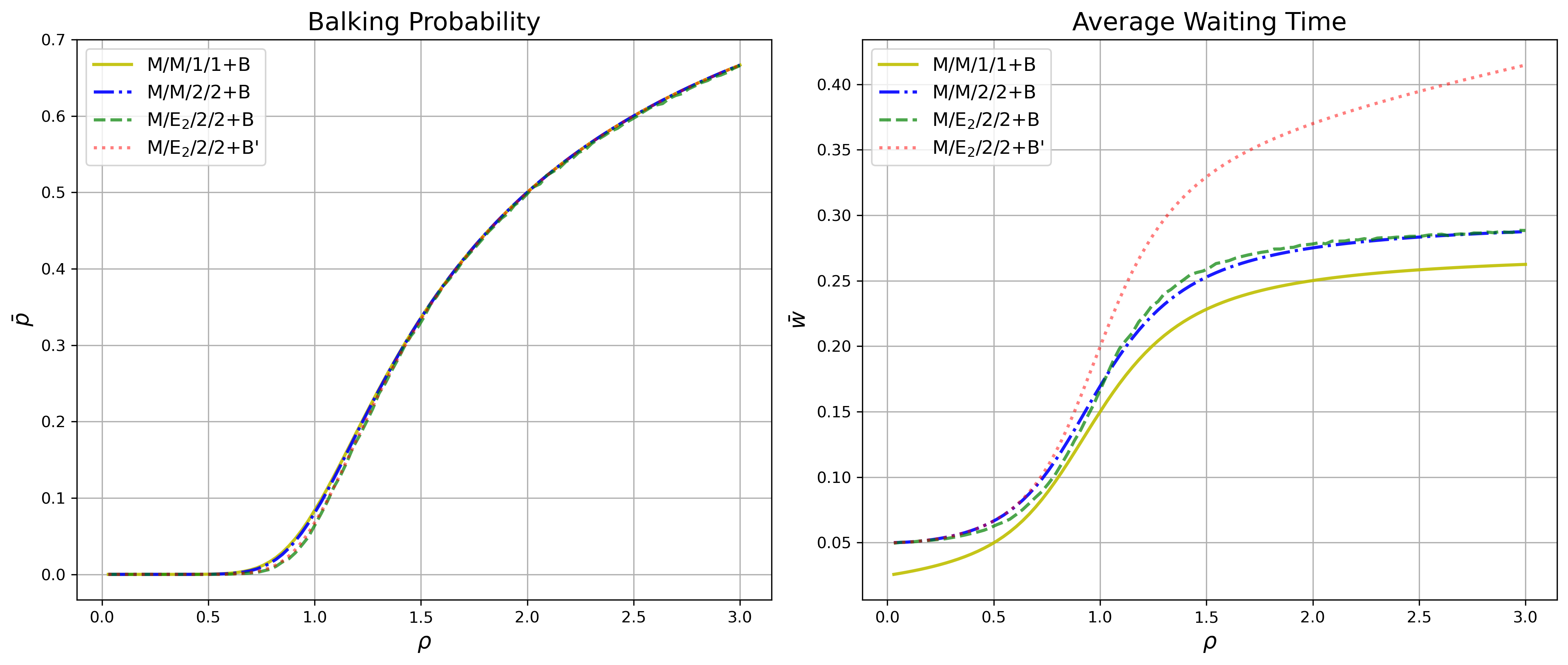}
    \caption{Queue performance comparison for $B=10$.}
    \label{fig:comparison_b10_s=2}
\end{figure}

Figures~\ref{fig:comparison_b0_s=2} and~\ref{fig:comparison_b10_s=2} plot the balking probability $\BalkingProbability = \ProbabilityOfNUsers{\QueueingSystemSize}$ and the average waiting time $\WaitingTime$ of the four queues for $B = 0$ and $B = 10$.
The horizontal axis is the utilization rate of the system $\rho$, defined as the ratio of the arrival rate $\lambda$ to the total service rate of all servers combined (40 in this experiment), that is, $\rho = \lambda / 40$.
Since the balking probability and the average waiting time of the $M/E_2/2/2 + B$ queues are not available analytically, we run simulations with 100 evenly spaced values of $\rho$s to obtain the estimate.\footnote{The simulation code is adapted from \url{https://github.com/lorcan2440/Process-Simulation/blob/main/Queueing/main.py}} Simulation results are averaged over 30 independent runs to mitigate statistical variability.

\paragraph{\boldmath\textbf{$M/M/1/1+B$ vs $M/M/2/2+B$}}
In Figure~\ref{fig:comparison_b0_s=2} we observe a noticeable difference in the probability of balking and average waiting time of the two queues.
In Figure~\ref{fig:comparison_b10_s=2}, the balking probabilities of the two queues are nearly identical, possibly due to the large buffer size, while the difference in average waiting time remains non-negligible.
Additional experiments presented in Appendix~\ref{appendix:queues_comparison} show the performance differences between single-server and multi-server queues become even more pronounced when the number of servers, $s$, is large.
These results suggest that the performance metrics of the $M/M/1/1+B$ queue may fail to approximate those of the $M/M/s/2+B$ queue with the same total service rate when $s \geq 2$.

\paragraph{\boldmath\textbf{$M/M/2/2+B$ vs $M/E_r/2/2+B$}} 
Figures~\ref{fig:comparison_b0_s=2} and~\ref{fig:comparison_b10_s=2} reveal that the differences in both the balking probability and average waiting time between the two queues are small. This is likely because the exponential distribution is a special case of the Erlang distribution. We note that the Erlang-1 distribution, which corresponds to $r = 1$, coincides with the exponential distribution. In the setting considered in this study (i.e. $r = 2$ and $s = 2$), the difference between the evaluated metrics of the two queues tends to be small.

\paragraph{\boldmath\textbf{$M/E_r/2/2+B$ vs $M/M/2/2+B'$}} 
The approximation method of~\cite{Smith2008MGCkPM}, the $M/M/2/2+B'$ queue, produces very close values of the balking probability compared to the $M/E_r/2/2+B$ queue, as shown in Figures~\ref{fig:comparison_b0_s=2} and~\ref{fig:comparison_b10_s=2}. In terms of the average waiting time, while they also align closely in Figure~\ref{fig:comparison_b0_s=2}, a substantial difference is observed in Figure~\ref{fig:comparison_b10_s=2}.

The above results indicate that, for our case study, the $M/M/s/K$ queue can serve as a proxy for the $M/E_r/s/K$ queue, providing a sufficiently accurate approximation of both the balking probability and the average waiting time. In particular, in the setting considered in this study, the $M/M/s/K$ queue tends to give a more accurate approximation than the method of~\cite{Smith2008MGCkPM}. Moreover, the analytical expressions of the balking probability and the average waiting time of the $M/M/s/K$ queue are available, making them well-suited for integration into the optimization framework. 
Therefore, in the remainder of this work, unless stated otherwise, we focus on the $M/M/\NServers/\QueueingSystemSize$ queues.
In the next section, we explain our approaches to solve the lower-level problem~\eqref{eq:llp} under the $M/M/\NServers/\QueueingSystemSize$ queues.

\subsection{Approximation of lower-level with $M/M/\NServers/\QueueingSystemSize$ queue}\label{subsec:mmsk}

This section discusses our approach to approximately solve the lower-level problem~\eqref{eq:llp} assuming the facilities are modeled as the $M/M/s/K$ queues.
We begin with a very useful proposition, whose proof is provided in Appendix~\ref{appendix:convexity_proof}.
\begin{prop}
\label{prop:llp_convex}
For $M/M/\NServers/\QueueingSystemSize$ queue, problem~\eqref{eq:llp} is convex.
\end{prop}

The convexity of the objective function facilitates the approximation of~\eqref{eq:llp}.
Consider the following convex functions, which appear in the objective of~\eqref{eq:llp}:
\begin{alignat}{2}
f^{\mathrm{l}}(\LinearisationDummyVar) &= \LinearisationDummyVar \ln \LinearisationDummyVar, \label{func:f^l_LLP}\\
f^{\mathrm{w}}_{\FacilityIndex}(\LinearisationDummyVar) &= \int_0^{\LinearisationDummyVar} \WaitingTime_{\FacilityIndex}(q)\,dq, && \qquad \forall \FacilityIndex \in \AllFacilityIndexSet, \label{func:f^w_LLP}\\
f^{\mathrm{p}}_{\FacilityIndex}(\LinearisationDummyVar) &= \int_0^{\LinearisationDummyVar} \BalkingProbability_{\FacilityIndex}(q)\,dq, && \qquad \forall \FacilityIndex \in \AllFacilityIndexSet. \label{func:f^p_LLP}
\end{alignat}

Given a set of points $\{\hat{\DisaggregatedArrivalRate}_{\LinearisationIndex} : \LinearisationIndex = 1, \ldots, \LinearisationNIndices \}$ evenly spaced in the interval $(0, \max_{\DemandIndex \in \DemandIndexSet} \{\DemandVolume_\DemandIndex \}]$ and $\{\hat{\ArrivalRate}_{\LinearisationIndex} : \LinearisationIndex = 1, \ldots, \LinearisationNIndices \}$ evenly spaced in the interval $[0, \sum_{\DemandIndex \in \DemandIndexSet}\{\DemandVolume_\DemandIndex\} ]$, one can construct piecewise-linear under estimators of the above functions as
\begin{alignat*}{2}
\hat{f}^{\mathrm{l}}(\LinearisationDummyVar) &= \max_{\LinearisationIndex = 1, \ldots, \LinearisationNIndices}\left\{ f^{\mathrm{l}}(\LinearisationPointAssignment) + \nabla f^{\mathrm{l}}(\LinearisationPointAssignment) (\LinearisationDummyVar - \LinearisationPointAssignment)\right\}, && \\
\hat{f}^{\mathrm{w}}_{\FacilityIndex}(\LinearisationDummyVar) &= \max_{\LinearisationIndex = 1, \ldots, \LinearisationNIndices}\left\{ f^{\mathrm{w}}_{\FacilityIndex}(\LinearisationPointAggregatedAssignment) + \nabla f^{\mathrm{w}}_{\FacilityIndex}(\LinearisationPointAggregatedAssignment) (\LinearisationDummyVar - \LinearisationPointAggregatedAssignment)\right\}, && \qquad \forall \FacilityIndex \in \AllFacilityIndexSet, \\
\hat{f}^{\mathrm{p}}_{\FacilityIndex}(\LinearisationDummyVar) &= \max_{\LinearisationIndex = 1, \ldots, \LinearisationNIndices}\left\{ f^{\mathrm{p}}_{\FacilityIndex}(\LinearisationPointAggregatedAssignment) + \nabla f^{\mathrm{p}}_{\FacilityIndex}(\LinearisationPointAggregatedAssignment) (\LinearisationDummyVar - \LinearisationPointAggregatedAssignment)\right\}, && \qquad \forall \FacilityIndex \in \AllFacilityIndexSet.
\end{alignat*}
Now, consider the following problem where functions $f^{\mathrm{l}}/f^{\mathrm{w}}/f^{\mathrm{p}}$ are replaced with the corresponding piecewise-linear under estimators $\hat{f}^{\mathrm{l}}/\hat{f}^{\mathrm{w}}/\hat{f}^{\mathrm{p}}$
\begin{alignat}{2}
\label{eq:llp-lp-compact}
\min_{\DisaggregatedArrivalRate, \ArrivalRate} \ & \sum_{\DemandIndex \in \DemandIndexSet} \sum_{\FacilityIndex \in \AllFacilityIndexSet} \left(
\dfrac{1}{\theta} \hat{f}^{\mathrm{l}}(\DisaggregatedArrivalRate_{\DemandIndex \FacilityIndex}) + \DisaggregatedArrivalRate_{\DemandIndex \FacilityIndex} \TravelTime_{\DemandIndex \FacilityIndex} 
\right) 
+ \sum_{\FacilityIndex \in \AllFacilityIndexSet} \left( \alpha \hat{f}^{\mathrm{w}}(\ArrivalRate_{\FacilityIndex})
+ \beta \hat{f}^{\mathrm{p}}(\ArrivalRate_{\FacilityIndex}) \right) \\
\text{s.t.} \ & 
\sum_{\FacilityIndex \in \AllFacilityIndexSet} \DisaggregatedArrivalRate_{\DemandIndex \FacilityIndex} = \DemandVolume_{\DemandIndex}, &&\forall \DemandIndex \in \DemandIndexSet \notag, \\
& \ArrivalRate_{\FacilityIndex} = \sum_{\DemandIndex \in \DemandIndexSet} \Build_{\DemandIndex \FacilityIndex}, &&\forall \FacilityIndex \in \AllFacilityIndexSet, \notag \\
& \DisaggregatedArrivalRate_{\DemandIndex \FacilityIndex} \le \DemandVolume_{\DemandIndex} \Build_{\FacilityIndex}, && \forall \DemandIndex \in \DemandIndexSet, \FacilityIndex \in \FacilityIndexSet, \notag \\
& \DisaggregatedArrivalRate_{\DemandIndex \FacilityIndex} \geq 0,  &&\forall \DemandIndex \in \DemandIndexSet, \FacilityIndex \in \AllFacilityIndexSet, \notag
\end{alignat}
which can be formulated as the following linear program
\begin{alignat}{3}
\label{eq:llp-lin}
\tag{LLP-lin($\Build$)}
\min_{\substack{\DisaggregatedArrivalRate, \ArrivalRate,\\\LinearisedFunctionValue^{\mathrm{l}}, \LinearisedFunctionValue^{\mathrm{w}}, \LinearisedFunctionValue^{\mathrm{p}}}} \ 
& 
\sum_{\DemandIndex \in \DemandIndexSet} \sum_{\FacilityIndex \in \AllFacilityIndexSet} \left(
\dfrac{1}{\theta} \LinearisedFunctionValue^{\mathrm{l}}_{\DemandIndex \FacilityIndex} + \DisaggregatedArrivalRate_{\DemandIndex \FacilityIndex} \TravelTime_{\DemandIndex \FacilityIndex} 
\right) 
+ \sum_{\FacilityIndex \in \AllFacilityIndexSet} \left( \alpha \LinearisedFunctionValue^{\mathrm{w}}_{\FacilityIndex}
+ \beta \LinearisedFunctionValue^{\mathrm{p}}_{\FacilityIndex} \right)
\\
\text{s.t.} \ & 
\sum_{\FacilityIndex \in \AllFacilityIndexSet} \DisaggregatedArrivalRate_{\DemandIndex \FacilityIndex} = \DemandVolume_{\DemandIndex}, && \qquad \forall \DemandIndex \in \DemandIndexSet \notag, && \NSHint{(\gamma)} \\
& \ArrivalRate_{\FacilityIndex} = \sum_{\DemandIndex \in \DemandIndexSet} \DisaggregatedArrivalRate_{\DemandIndex \FacilityIndex}, && \qquad \forall \FacilityIndex \in \AllFacilityIndexSet, \notag && \NSHint{(\delta)} \\
& \DisaggregatedArrivalRate_{\DemandIndex \FacilityIndex} \le \DemandVolume_{\DemandIndex} \Build_{\FacilityIndex}, && \qquad \forall \DemandIndex \in \DemandIndexSet, \FacilityIndex \in \FacilityIndexSet, \notag && \NSHint{(\xi)} \\
& \LinearisedFunctionValue^{\mathrm{l}}_{\DemandIndex \FacilityIndex} \ge f^{\mathrm{l}}(\LinearisationPointAssignment) + \nabla f^{\mathrm{l}}(\LinearisationPointAssignment) (\DisaggregatedArrivalRate_{\DemandIndex \FacilityIndex} - \LinearisationPointAssignment), && \qquad \forall \DemandIndex \in \DemandIndexSet, \FacilityIndex \in \AllFacilityIndexSet, \LinearisationIndex = 1, \ldots, \LinearisationNIndices, \notag && \NSHint{(\nu^{\mathrm{l}})} \\
& \LinearisedFunctionValue^{\mathrm{w}}_{\FacilityIndex} \ge f^{\mathrm{w}}_{\FacilityIndex}(\LinearisationPointAggregatedAssignment) + \nabla f^{\mathrm{w}}_{\FacilityIndex}(\LinearisationPointAggregatedAssignment) (\ArrivalRate_{\FacilityIndex} - \LinearisationPointAggregatedAssignment), && \qquad \forall \FacilityIndex \in \AllFacilityIndexSet, \LinearisationIndex = 1, \ldots, \LinearisationNIndices, \notag && \NSHint{(\nu^{\mathrm{w}})} \\
& \LinearisedFunctionValue^{\mathrm{p}}_{\FacilityIndex} \ge f^{\mathrm{p}}_{\FacilityIndex}(\LinearisationPointAggregatedAssignment) + \nabla f^{\mathrm{p}}_{\FacilityIndex}(\LinearisationPointAggregatedAssignment) (\ArrivalRate_{\FacilityIndex} - \LinearisationPointAggregatedAssignment), && \qquad \forall \FacilityIndex \in \AllFacilityIndexSet, \LinearisationIndex = 1, \ldots, \LinearisationNIndices, \notag && \NSHint{(\nu^{\mathrm{p}})} \\
& \DisaggregatedArrivalRate_{\DemandIndex \FacilityIndex} \geq 0,  && \qquad \forall \DemandIndex \in \DemandIndexSet, \FacilityIndex \in \AllFacilityIndexSet. \notag
\end{alignat}

To reduce clutter, we will write~\eqref{eq:llp-lin} as
\begin{subequations}
\label{eq:llp-lin-compact}
\begin{alignat}{2}
\min_{\DisaggregatedArrivalRate, \ArrivalRate, \LinearisedFunctionValue} \ 
& 
c_{\DisaggregatedArrivalRate}^{\top} \DisaggregatedArrivalRate
+ c_{\ArrivalRate}^{\top} \ArrivalRate
+ c_{\LinearisedFunctionValue}^{\top} \LinearisedFunctionValue
\\
\text{s.t.} \ & 
A_{\DisaggregatedArrivalRate} \DisaggregatedArrivalRate
+ A_{\ArrivalRate} \ArrivalRate
+ A_{\LinearisedFunctionValue} \LinearisedFunctionValue
\ge b - B \Build,
\label{eq:llp-lin-compact-constraint}
\end{alignat}
\end{subequations}
where $\LinearisedFunctionValue = (\LinearisedFunctionValue^{\mathrm{l}}, \LinearisedFunctionValue^{\mathrm{w}}, \LinearisedFunctionValue^{\mathrm{p}})$ and $c$, $A$, $b$, $B$ are vectors/matrices of appropriate dimensions.
We note that we included the bound constraint $\DisaggregatedArrivalRate \ge 0$ in the newly defined constraints~\eqref{eq:llp-lin-compact-constraint}.

In the next section, we describe the formulations of the FLP and our approaches, in which we use the linear approximation~\eqref{eq:llp-lin}.

\section{Bilevel formulation of facility location problem}\label{sec:bilevel_formulation_of_facility_location_problem}

Assuming the underlying queueus are $M/M/\NServers/\QueueingSystemSize$ queues, the bilevel facility location problem~\eqref{eq:abstract_flp} can be written as
\begin{alignat}{2}
\label{eq:flp}
\tag{FLP}
  \max_{\Build, \DisaggregatedArrivalRate, \ArrivalRate} \ & 
  \sum_{\FacilityIndex \in \FacilityIndexSet} \ArrivalRate_{\FacilityIndex} (1 - \BalkingProbability_{\FacilityIndex}(\ArrivalRate_{\FacilityIndex}))
  \\
  \text{s.t.}\ 
  & 
  \Build \in \BuildSet,
  (\DisaggregatedArrivalRate, \ArrivalRate) \in \mathcal{S}(\text{\ref{eq:llp}}), \notag
\end{alignat}
where $\ArrivalRate_{\FacilityIndex} (1 - \BalkingProbability_{\FacilityIndex}(\ArrivalRate_{\FacilityIndex}))$ is the throughput of the facility (queue) $\FacilityIndex$ and $\mathcal{S}(\text{\ref{eq:llp}})$ is the solution set for~\eqref{eq:llp}. To solve the~\eqref{eq:flp}, we follow the approaches taken by~\cite{dan19}, which are illustrated in the following two subsections. While \cite{dan19} consider $M/M/1/K$ queues, our contribution lies in embedding the more general $M/M/s/K$ queueing system into this bilevel optimization framework, allowing for more realistic modeling of facilities with multiple charging outlets.

\subsection{Linearization approach}\label{subsec:linearization_approach}

The first approach, referred to as the \emph{linearization approach}, solves~\eqref{eq:flp} approximately by replacing the lower-level problem~\eqref{eq:llp} with its linear approximation~\eqref{eq:llp-lin} and linearizing the upper-level objective.

By replacing the lower-level problem, we obtain
\begin{alignat}{2}
\label{eq:flp-lin}
\tag{FLP-lin}
  \max_{\Build, \DisaggregatedArrivalRate, \ArrivalRate} \ & 
  \sum_{\FacilityIndex \in \FacilityIndexSet} \ArrivalRate_{\FacilityIndex} (1 - \BalkingProbability_{\FacilityIndex}(\ArrivalRate_{\FacilityIndex}))
  \\
  \text{s.t.}\ 
  & 
  \Build \in \BuildSet,
  (\DisaggregatedArrivalRate, \ArrivalRate) \in \mathcal{S}(\text{\ref{eq:llp-lin}}). \notag
\end{alignat}
We can reformulate this bilevel programming model as a single-level problem by replacing the lower-level problem with its Karush–Kuhn–Tucker (KKT) conditions as follows:
\begin{alignat*}{2}
\max_{\substack{\Build, \DisaggregatedArrivalRate, \ArrivalRate, \\\LinearisedFunctionValue, \pi}}
\ & 
  \sum_{\FacilityIndex \in \FacilityIndexSet} \ArrivalRate_{\FacilityIndex} (1 - \BalkingProbability_{\FacilityIndex}(\ArrivalRate_{\FacilityIndex}))
  \\
  \text{s.t.}\ 
&  A_{\DisaggregatedArrivalRate}\DisaggregatedArrivalRate
+ A_{\ArrivalRate} \ArrivalRate
+ A_{\LinearisedFunctionValue} \LinearisedFunctionValue
\ge b - B \Build, \\
& c_{\DisaggregatedArrivalRate}^{\top} \DisaggregatedArrivalRate
+ c_{\ArrivalRate}^{\top} \ArrivalRate
+ c_{\LinearisedFunctionValue}^{\top} \LinearisedFunctionValue
\le
(b - B \Build)^{\top} \pi, \\
& A_{\DisaggregatedArrivalRate}^{\top} \pi = c_{\DisaggregatedArrivalRate}, \\
& A_{\ArrivalRate}^{\top} \pi = c_{\ArrivalRate}, \\
& A_{\LinearisedFunctionValue}^{\top} \pi = c_{\LinearisedFunctionValue}, \\
  & 
  \Build \in \BuildSet, \pi \ge 0. \notag
\end{alignat*}
There is a bilinear term $\pi^{\top} B \Build$.
One can ``linearize'' this term by exploiting the fact that $\Build$ is binary.
More specifically, one can use the indicator constraint and write
\begin{alignat}{2}
\label{eq:linearisation_approach_with_nonlinear_objective}
\max_{\substack{\Build, \DisaggregatedArrivalRate, \ArrivalRate, \\\LinearisedFunctionValue, \pi, \Pi}}
\ & 
  \sum_{\FacilityIndex \in \FacilityIndexSet} \ArrivalRate_{\FacilityIndex} (1 - \BalkingProbability_{\FacilityIndex}(\ArrivalRate_{\FacilityIndex}))
  \\
  \text{s.t.}\ 
&  A_{\DisaggregatedArrivalRate}\DisaggregatedArrivalRate
+ A_{\ArrivalRate}^{\top} \ArrivalRate
+ A_{\LinearisedFunctionValue} \LinearisedFunctionValue
\ge b - B \Build, \notag \\
& c_{\DisaggregatedArrivalRate}^{\top} \DisaggregatedArrivalRate
+ c_{\ArrivalRate}^{\top} \ArrivalRate
+ c_{\LinearisedFunctionValue}^{\top} \LinearisedFunctionValue
\le
b^{\top} \pi - \mathrm{tr}(B^{\top} \Pi), \notag \\
& A_{\DisaggregatedArrivalRate}^{\top} \pi = c_{\DisaggregatedArrivalRate}, \notag \\
& A_{\ArrivalRate}^{\top} \pi = c_{\ArrivalRate}, \notag \\
& A_{\LinearisedFunctionValue}^{\top} \pi = c_{\LinearisedFunctionValue}, \notag \\
&
\Build_{\FacilityIndex} = 1
\Longrightarrow
\Pi_{\ConstraintIndex \FacilityIndex} = \pi_{\ConstraintIndex},
&& \ \forall \ConstraintIndex = 1, \ldots, \NConstraints, \FacilityIndex \in \FacilityIndexSet, \notag
\notag \\
&
\Build_{\FacilityIndex} = 0
\Longrightarrow
\Pi_{\ConstraintIndex \FacilityIndex} = 0,
&& \ \forall \ConstraintIndex = 1, \ldots, \NConstraints, \FacilityIndex \in \FacilityIndexSet, \notag
\notag \\
  & 
  \Build \in \BuildSet, \pi \ge 0. \notag
\end{alignat}
where $\NConstraints$ is the number of constraints in~\eqref{eq:llp-lin-compact-constraint}.

Next, we approximate the objective with a piecewise-linear approximation \citep{DAMBROSIO201039}.
Let
$$
f^{\mathrm{t}}_{\FacilityIndex}(\LinearisationDummyVar) = \LinearisationDummyVar (1 - \BalkingProbability_{\FacilityIndex}(\LinearisationDummyVar)),
\qquad \forall \FacilityIndex \in \FacilityIndexSet.
$$
Suppose $\{\hat{\ArrivalRate}_{\LinearisationIndex} : \LinearisationIndex = 1, \ldots, \LinearisationNIndices \}$ is sorted in increasing order, and let
$$
\hat{f}^{\mathrm{t}}_{\FacilityIndex}(\LinearisationDummyVar) = 
\begin{cases}
\alpha f^{\mathrm{t}}_{\FacilityIndex}(\hat{\ArrivalRate}_{\LinearisationIndex})
+
(1 - \alpha) f^{\mathrm{t}}_{\FacilityIndex}(\hat{\ArrivalRate}_{\LinearisationIndex + 1}), &
\text{ if $\LinearisationDummyVar = \alpha \hat{\ArrivalRate}_{\LinearisationIndex} + (1 - \alpha) \hat{\ArrivalRate}_{\LinearisationIndex + 1}$ for some $\LinearisationIndex$ and $\alpha \in [0, 1]$}, \\
\infty, &
\text{ otherwise,}
\end{cases}
$$
for all $\FacilityIndex \in \FacilityIndexSet$.
We replace $f^{\mathrm{t}}_{\FacilityIndex}$ in the objective in~\eqref{eq:linearisation_approach_with_nonlinear_objective} with its approximation $\hat{f}^{\mathrm{t}}_{\FacilityIndex}$.
The resulting optimization problem can be written as
\begin{alignat}{2}
\label{eq:linearisation_approach_with_linear_objective}
\max_{\substack{\Build, \DisaggregatedArrivalRate, \ArrivalRate, \\\LinearisedFunctionValue, \pi, \Pi, \alpha}}
\ & 
  \sum_{\FacilityIndex \in \FacilityIndexSet}
  \sum_{\LinearisationIndex = 1}^{\LinearisationNIndices}
  \alpha_{\FacilityIndex \LinearisationIndex} f^{\mathrm{t}}_{\FacilityIndex}(\hat{\ArrivalRate}_{\LinearisationIndex})
  \\
  \text{s.t.}\ 
  & \ArrivalRate_{\FacilityIndex} = \sum_{\LinearisationIndex = 1}^{\LinearisationNIndices}
  \alpha_{\FacilityIndex \LinearisationIndex} \hat{\ArrivalRate}_{\LinearisationIndex}, 
  && \ \forall \FacilityIndex \in \FacilityIndexSet, \notag
  \\
  & \sum_{\LinearisationIndex = 1}^{\LinearisationNIndices}
  \alpha_{\FacilityIndex \LinearisationIndex} = 1,
  && \ \forall \FacilityIndex \in \FacilityIndexSet, \notag
  \\
  & \mathrm{SOS2}(\{\alpha_{\FacilityIndex \LinearisationIndex} : \LinearisationIndex = 1, \ldots, \LinearisationNIndices\}), 
  && \ \forall \FacilityIndex \in \FacilityIndexSet, \notag
  \\
&  A_{\DisaggregatedArrivalRate}\DisaggregatedArrivalRate
+ A_{\ArrivalRate}^{\top} \ArrivalRate
+ A_{\LinearisedFunctionValue} \LinearisedFunctionValue
\ge b - B \Build, \notag \\
& c_{\DisaggregatedArrivalRate}^{\top} \DisaggregatedArrivalRate
+ c_{\ArrivalRate}^{\top} \ArrivalRate
+ c_{\LinearisedFunctionValue}^{\top} \LinearisedFunctionValue
\le
b^{\top} \pi - \mathrm{tr}(B^{\top} \Pi), \notag \\
& A_{\DisaggregatedArrivalRate}^{\top} \pi = c_{\DisaggregatedArrivalRate}, \notag \\
& A_{\ArrivalRate}^{\top} \pi = c_{\ArrivalRate}, \notag \\
& A_{\LinearisedFunctionValue}^{\top} \pi = c_{\LinearisedFunctionValue}, \notag \\
&
\Build_{\FacilityIndex} = 1
\Longrightarrow
\Pi_{\ConstraintIndex \FacilityIndex} = \pi_{\ConstraintIndex},
&& \ \forall \ConstraintIndex = 1, \ldots, \NConstraints, \FacilityIndex \in \FacilityIndexSet, \notag
\notag \\
&
\Build_{\FacilityIndex} = 0
\Longrightarrow
\Pi_{\ConstraintIndex \FacilityIndex} = 0,
&& \ \forall \ConstraintIndex = 1, \ldots, \NConstraints, \FacilityIndex \in \FacilityIndexSet, \notag
\notag \\
  & 
  \Build \in \BuildSet, \pi \ge 0, \alpha \ge 0. \notag
\end{alignat}

\subsection{Surrogate-based heuristic}\label{subsec:surrogate_based_heuristic}

The second approach is a heuristic that quickly provides a bilevel-feasible solution.
In this approach, we solve the following problem
\begin{alignat}{2}
\label{eq:surrogate_based_heuristic_bilevel_formulation}
  \max_{\Build, \DisaggregatedArrivalRate, \ArrivalRate} \ & 
  -(c_{\DisaggregatedArrivalRate}^{\top} \DisaggregatedArrivalRate
+ c_{\ArrivalRate}^{\top} \ArrivalRate
+ c_{\LinearisedFunctionValue}^{\top} \LinearisedFunctionValue)
  \\
  \text{s.t.}\ 
  & 
  \Build \in \BuildSet,
  (\DisaggregatedArrivalRate, \ArrivalRate) \in \mathcal{S}(\text{\ref{eq:llp-lin}}), \notag
\end{alignat}
which is the same as problem~\eqref{eq:flp-lin} except that the objective is replaced with that of the lower-level problem~\eqref{eq:llp-lin}.
Since the upper-level and the lower-level objectives coincide, it can be shown that this problem is equivalent to
\begin{alignat*}{2}
\min_{\Build, \DisaggregatedArrivalRate, \ArrivalRate, \LinearisedFunctionValue} \ 
& 
c_{\DisaggregatedArrivalRate}^{\top} \DisaggregatedArrivalRate
+ c_{\ArrivalRate}^{\top} \ArrivalRate
+ c_{\LinearisedFunctionValue}^{\top} \LinearisedFunctionValue
\\
\text{s.t.} \ & 
A_{\DisaggregatedArrivalRate} \DisaggregatedArrivalRate
+ A_{\ArrivalRate} \ArrivalRate
+ A_{\LinearisedFunctionValue} \LinearisedFunctionValue
\ge b - B \Build, \notag \\
& \Build \in \BuildSet.
\end{alignat*}
We will refer to this approach as the \emph{surrogate-based heuristic}.

\section{Case study}\label{sec:casestudy}

This section applies the previously introduced bilevel optimization model to a real-world setting involving the tactical placement of EV charging stations in a residential area of the city of Montreal, the \emph{Le Plateau-Mont-Royal}. In this context, facilities refer to EV charging stations (sites), servers within a facility correspond to outlets at a station, and demand represents EV users.

Our case study is divided into two main parts, corresponding to the two levels of the model. In Section~\ref{subsec:casestudy_lowerlevel}, we focus on the lower-level problem modeling the user equilibrium flows based on their disutilities. We describe our case study, including how we obtain the set of users, their demands and the set of sites. We also explain how we calibrate the parameters $\alpha, \beta$ and $\theta^{-1}$ appearing in the user choice model (recall Section~\ref{subsec:userchoice}) using real-world data from existent charging stations. In addition, we investigate the sensitivity of our linear approximation~\eqref{eq:llp-lin} to the number of linearization points $\LinearisationNIndices$ used to obtain it (recall Section~\ref{subsec:mmsk}). In Section~\ref{subsec:casestudy_upperlevel}, we move on to apply our bilevel optimization model, where the aim is to determine the optimal upper-level decisions, i.e. the charging station locations. This study considers different scenarios of interest, including varying buffer sizes, user choice parameters, budget constraint for new stations and competition.

\subsection{Lower-level problem}\label{subsec:casestudy_lowerlevel}

In this section, we describe the setup of the data and instance generation process that underpins our case study, including elements that carry over to the subsequent analysis in Section~\ref{subsec:casestudy_upperlevel}. Since our available data reports completed charging sessions rather than user arrivals, we cannot directly observe queueing behavior such as balking or waiting times. As a result, we calibrate the user choice model parameters via a grid search to best align predicted and observed charging activity. As part of this calibration, we also identify appropriate values for the number of linearization points used in our approximation of the lower-level problem, which will be maintained throughout the remainder of the case study. 

\subsubsection{Data and instance generation}

\begin{figure}[h!]
    \centering
    \includegraphics[scale=0.25]{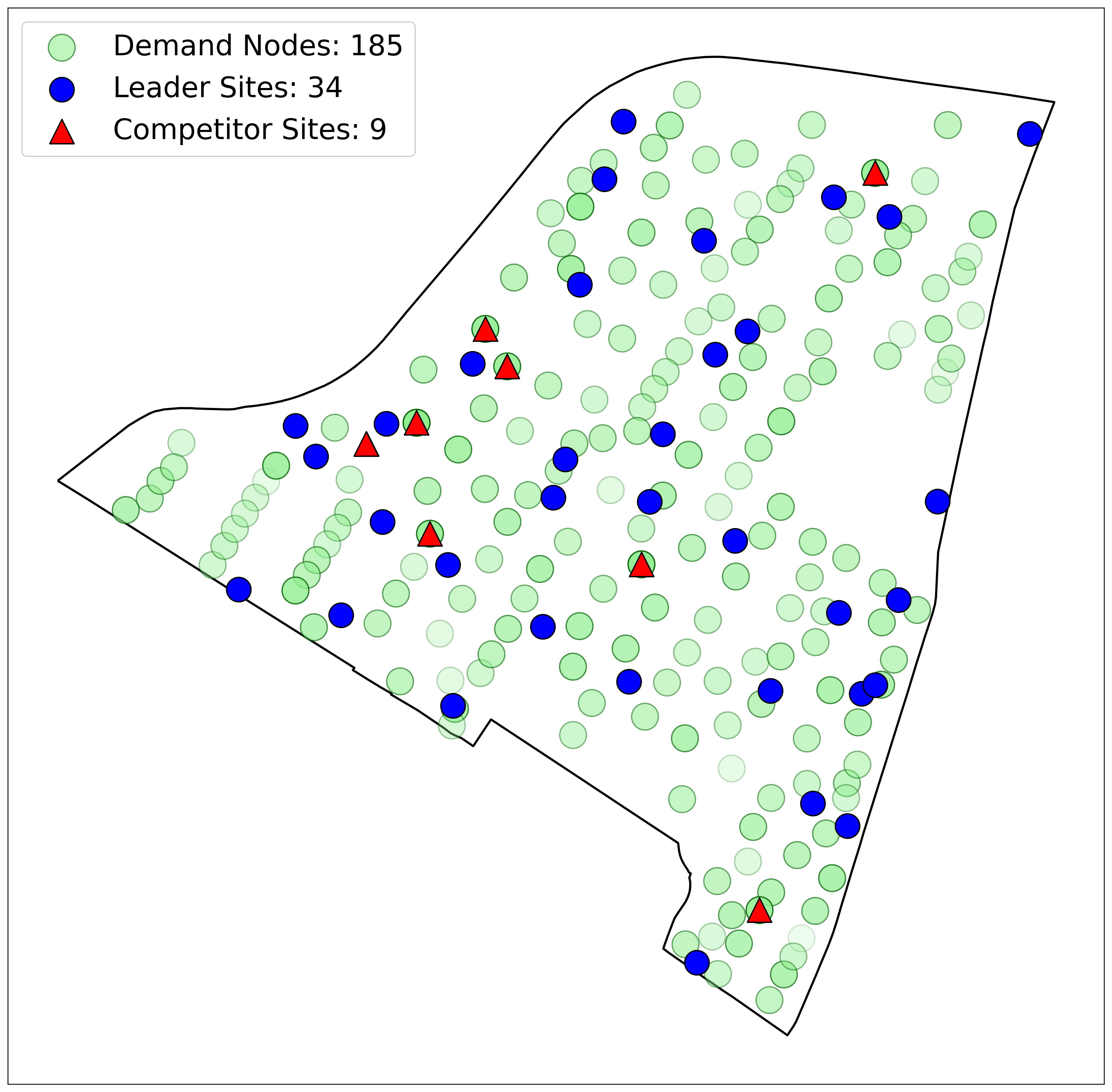}
    \caption{Le Plateau-Mont-Royal network displaying the locations of the demand nodes, existing charging stations operated by the leader and competitors' sites.}
    \label{fig:original_map_comp}
\end{figure}

Our case study relies on real-world data and modeling assumptions designed to reflect realistic urban charging infrastructure conditions. Next, we describe how we use the available data to construct lower-level instances.

Real-world population data from the 2021 Census is obtained from \cite{StatisticsCanada2021}. This data provides 185 population centroids in the residential borough of Le Plateau-Mont-Royal in Montreal, which we use as the demand set $I$; see Figure~\ref{fig:original_map_comp}. The green demand nodes in the figure represent these centroids, with color saturation proportional to population size—darker shades indicate higher population density. This area is selected because it is one of the most densely populated urban neighborhoods in the city, making it a relevant context where congestion plays an important role. To set the demands from $I$, we base them on charging session data from the existing public charging network in the study area. In particular, we use the number of daily charging sessions recorded between January 1, 2023, and January 10, 2024. Based on our statistical analysis, the median number of daily charging sessions corresponds to approximately 0.2\% of the population, while the 0.99 quantile (i.e., the busiest days) accounts for about 0.3\% of the population. Hence, to account for potential demand associated with implicit behaviors such as balking and reneging (for which data are not available), we consider a demand factor of 0.3\%. More concretely, each population centroid $i$ has a demand $d_i$ equal to its populations multiplied by a uniform rate of 0.003.


The network from our dataset consists of 36 existing public EV charging stations, 34 stations are operated by \textit{Circuit Électrique} while 2 stations are by \textit{FLO}. We assume Circuit Électrique is the leader, and FLO is the competitor in this study. Figure~\ref{fig:original_map_comp} shows 34 blue circles representing the leader sites. The 2 triangles that do not overlap with green demand nodes correspond to effective competitor sites—these 2 are positioned very close to each other, which is why only 1 is visually distinguishable on the map. The remaining 7 overlapping triangles indicate additional competitor stations that will be introduced later in Section~\ref{subsec:casestudy_upperlevel}. Among the 36 existing sites, there is one Level-3 (L3) single-outlet site and one Level-2 (L2) single-outlet site; the remaining sites are all L2 with two outlets each. L3 chargers provide significantly higher charging power than L2 chargers, resulting in shorter charging durations and improved site throughput.
The service times of charging sessions for L2 and L3 stations were fitted to Erlang-$r$ distributions, yielding a service rate $\mu$ of 9 and 58 users per day, respectively. Although the actual service time distribution can be better captured by an $M/E_r/s/K$ queue, each site is modeled as an $M/M/s/K$ queue with the same capacity parameters $s$ and $K$ as a surrogate approximation, motivated by the analysis of queueing systems (recall Section~\ref{subsec:queues_comparison}). 
For each station $\FacilityIndex \in \AllFacilityIndexSet$, the buffer size is uniformly set across all stations, testing two values from the set ${0, 2}$ to represent typical urban parking constraints with minimal waiting space. Travel times $t_{ij}$ between each centroid~$i$ and charging site $j$ were computed based on the Euclidean distance (under coordinate system EPSG:3347) and the driving speed of 30~km/h. 

In our experiments tackling the bilevel optimization, we consider the values of the disutility parameters $(\alpha, \beta, \theta)$ to be in the set $\{ (0,10,0), (10,10,0)\}$. The linear approximation used to solve the lower-level model is conducted with $\LinearisationNIndices = 100$ equidistant points for both the disaggregated arrival rate $\DisaggregatedArrivalRate$ and the aggregated arrival rate $\ArrivalRate$ (recall Section~\ref{subsec:mmsk}). The justification for the disutility parameters and the number of linearization points is explained in the next section.

\subsubsection{Calibration}\label{subsubsec:calibration} 
In this section, we aim to calibrate~\eqref{eq:llp-lin}, namely we aim to find reasonable values for $(\alpha, \beta, \theta)$ and the number of linearization points $\LinearisationNIndices$.

Considering the network of 36 existing public EV charging stations, we calibrate the user disutility parameters in the triplet $(\alpha, \beta, \theta)$ via grid search: $\alpha$ and $\beta$ were varied from 0 to 50 in increments of 10, and $\theta^{-1}$ was varied from 0 to 5 in increments of 1. To evaluate the accuracy of each parameter combination, the Kullback–Leibler (KL) divergence \citep{kullback1951information} is computed between the normalized recorded number of daily charging sessions and the normalized predicted number of daily arrivals, coming from~\eqref{eq:llp-lin}. The comparison is performed over a 30-day period (December 12, 2023, to January 10, 2024), using the most recent available data. This window is selected to best capture current demand patterns, given that charging demand is non-stationary over time. It is important to note that we lack data on queues formed at stations or on users who arrived when a site was already occupied. This justifies our simple calibration procedure.

Our calibration results for the top 60 performing triplets, under two buffer size settings (zero and two), are presented in Table~\ref{tab:kl-top60-b=0} and Table~\ref{tab:kl-top60-b=2}, respectively. For both cases of buffer sizes, $(0,10,0)$ is the triplet that achieves the best score (i.e., the lowest) of KL divergence. This case means $\alpha = 0$ and $\theta^{-1} = 0$ (i.e., $\theta \to \infty$), indicating that users exhibit negligible sensitivity to average waiting time and stochasticity. This result suggests that users behave in accordance with Wardrop equilibrium conditions, where $1/\theta = 0$. In this context, EV users appear to base their decisions solely on travel time and the likelihood of balking, completely avoiding sites with severe congestion. This behavior is intuitively reasonable, particularly in residential settings where extended waiting times and limited parking availability discourages users from queueing. 

\begin{table}[h!]
\centering
\caption{KL divergence results for buffer size $B=0$ (ranks 1–60)}
\label{tab:kl-top60-b=0}

\noindent
\begin{minipage}{0.32\textwidth}
\begin{tabular}{rrrrr}
\toprule
Idx & $\alpha$ & $\beta$ & $\theta^{-1}$ & KL \\
\midrule
1  & 0  & 10 & 0 & 0.0502 \\
2  & 10 & 10 & 0 & 0.0536 \\
3  & 0  & 10 & 2 & 0.0568 \\
4  & 10 & 10 & 2 & 0.0576 \\
5  & 0  & 20 & 2 & 0.0577 \\
6  & 10 & 20 & 5 & 0.0586 \\
7  & 0  & 20 & 4 & 0.0587 \\
8  & 10 & 20 & 4 & 0.0588 \\
9  & 0  & 20 & 5 & 0.0589 \\
10 & 10 & 20 & 2 & 0.0589 \\
11 & 10 & 30 & 5 & 0.0591 \\
12 & 0  & 30 & 5 & 0.0591 \\
13 & 0  & 30 & 2 & 0.0592 \\
14 & 20 & 20 & 5 & 0.0595 \\
15 & 20 & 20 & 4 & 0.0597 \\
16 & 0  & 20 & 3 & 0.0597 \\
17 & 0  & 30 & 4 & 0.0597 \\
18 & 10 & 30 & 4 & 0.0598 \\
19 & 0  & 40 & 5 & 0.0599 \\
20 & 10 & 40 & 5 & 0.0599 \\
\bottomrule
\end{tabular}
\end{minipage}
\hfill
\begin{minipage}{0.32\textwidth}
\begin{tabular}{rrrrr}
\toprule
Idx & $\alpha$ & $\beta$ & $\theta^{-1}$ & KL \\
\midrule
21 & 30 & 20 & 5 & 0.0599 \\
22 & 20 & 10 & 0 & 0.0600 \\
23 & 10 & 30 & 2 & 0.0601 \\
24 & 20 & 30 & 5 & 0.0601 \\
25 & 0  & 40 & 4 & 0.0601 \\
26 & 10 & 40 & 4 & 0.0602 \\
27 & 30 & 30 & 5 & 0.0602 \\
28 & 10 & 20 & 3 & 0.0603 \\
29 & 0  & 50 & 5 & 0.0604 \\
30 & 10 & 50 & 5 & 0.0604 \\
31 & 10 & 10 & 5 & 0.0606 \\
32 & 0  & 10 & 3 & 0.0607 \\
33 & 0  & 30 & 3 & 0.0607 \\
34 & 0  & 10 & 4 & 0.0608 \\
35 & 20 & 30 & 4 & 0.0609 \\
36 & 20 & 40 & 5 & 0.0609 \\
37 & 20 & 40 & 4 & 0.0609 \\
38 & 30 & 40 & 5 & 0.0610 \\
39 & 10 & 10 & 4 & 0.0611 \\
40 & 0  & 40 & 2 & 0.0612 \\
\bottomrule
\end{tabular}
\end{minipage}
\hfill
\begin{minipage}{0.32\textwidth}
\begin{tabular}{rrrrr}
\toprule
Idx & $\alpha$ & $\beta$ & $\theta^{-1}$ & KL \\
\midrule
41 & 0  & 10 & 5 & 0.0612 \\
42 & 20 & 30 & 2 & 0.0612 \\
43 & 30 & 40 & 4 & 0.0612 \\
44 & 10 & 30 & 3 & 0.0612 \\
45 & 10 & 10 & 3 & 0.0613 \\
46 & 20 & 50 & 5 & 0.0615 \\
47 & 20 & 30 & 3 & 0.0616 \\
48 & 30 & 50 & 5 & 0.0616 \\
49 & 20 & 20 & 3 & 0.0616 \\
50 & 20 & 20 & 2 & 0.0617 \\
51 & 40 & 40 & 5 & 0.0619 \\
52 & 30 & 30 & 4 & 0.0619 \\
53 & 40 & 50 & 5 & 0.0619 \\
54 & 20 & 10 & 2 & 0.0620 \\
55 & 20 & 10 & 5 & 0.0620 \\
56 & 30 & 20 & 4 & 0.0621 \\
57 & 40 & 30 & 5 & 0.0621 \\
58 & 40 & 20 & 5 & 0.0622 \\
59 & 0  & 40 & 3 & 0.0622 \\
60 & 0  & 50 & 4 & 0.0622 \\
\bottomrule
\end{tabular}
\end{minipage}

\end{table}

\begin{table}[h!]
\centering
\caption{KL divergence results for buffer size = 2 (ranks 1–60)}
\label{tab:kl-top60-b=2}
\noindent
\begin{minipage}{0.32\textwidth}
\begin{tabular}{rrrrr}
\toprule
Idx & $\alpha$ & $\beta$ & $\theta^{-1}$ & KL \\
\midrule
1  & 0  & 10 & 0 & 0.0327 \\
2  & 0  & 20 & 0 & 0.0437 \\
3  & 10 & 10 & 0 & 0.0498 \\
4  & 0  & 30 & 0 & 0.0516 \\
5  & 0  & 40 & 0 & 0.0529 \\
6  & 10 & 20 & 0 & 0.0554 \\
7  & 10 & 30 & 0 & 0.0558 \\
8  & 10 & 40 & 0 & 0.0562 \\
9  & 0  & 50 & 0 & 0.0572 \\
10 & 0  & 30 & 2 & 0.0587 \\
11 & 0  & 20 & 2 & 0.0590 \\
12 & 0  & 40 & 5 & 0.0597 \\
13 & 0  & 10 & 1 & 0.0598 \\
14 & 0  & 50 & 5 & 0.0600 \\
15 & 0  & 40 & 3 & 0.0604 \\
16 & 0  & 50 & 4 & 0.0605 \\
17 & 0  & 30 & 5 & 0.0605 \\
18 & 0  & 40 & 4 & 0.0606 \\
19 & 10 & 40 & 5 & 0.0607 \\
20 & 10 & 50 & 5 & 0.0609 \\
\bottomrule
\end{tabular}
\end{minipage}
\hfill
\begin{minipage}{0.32\textwidth}
\begin{tabular}{rrrrr}
\toprule
Idx & $\alpha$ & $\beta$ & $\theta^{-1}$ & KL \\
\midrule
21 & 0  & 30 & 4 & 0.0610 \\
22 & 0  & 50 & 3 & 0.0612 \\
23 & 20 & 40 & 0 & 0.0614 \\
24 & 0  & 30 & 3 & 0.0615 \\
25 & 10 & 30 & 5 & 0.0616 \\
26 & 0  & 40 & 2 & 0.0616 \\
27 & 0  & 10 & 2 & 0.0618 \\
28 & 0  & 50 & 2 & 0.0619 \\
29 & 10 & 50 & 0 & 0.0620 \\
30 & 0  & 20 & 4 & 0.0622 \\
31 & 0  & 20 & 3 & 0.0624 \\
32 & 10 & 40 & 4 & 0.0625 \\
33 & 0  & 20 & 5 & 0.0627 \\
34 & 10 & 30 & 4 & 0.0628 \\
35 & 10 & 40 & 3 & 0.0628 \\
36 & 20 & 50 & 0 & 0.0629 \\
37 & 20 & 50 & 5 & 0.0629 \\
38 & 10 & 50 & 4 & 0.0631 \\
39 & 10 & 30 & 3 & 0.0632 \\
40 & 10 & 20 & 5 & 0.0633 \\
\bottomrule
\end{tabular}
\end{minipage}
\hfill
\begin{minipage}{0.32\textwidth}
\begin{tabular}{rrrrr}
\toprule
Idx & $\alpha$ & $\beta$ & $\theta^{-1}$ & KL \\
\midrule
41 & 20 & 40 & 5 & 0.0633 \\
42 & 10 & 20 & 4 & 0.0634 \\
43 & 10 & 50 & 3 & 0.0635 \\
44 & 20 & 30 & 5 & 0.0637 \\
45 & 0  & 30 & 1 & 0.0642 \\
46 & 0  & 20 & 1 & 0.0646 \\
47 & 20 & 30 & 0 & 0.0649 \\
48 & 20 & 20 & 5 & 0.0649 \\
49 & 0  & 10 & 3 & 0.0651 \\
50 & 10 & 20 & 2 & 0.0652 \\
51 & 10 & 30 & 2 & 0.0653 \\
52 & 10 & 0  & 2 & 0.0654 \\
53 & 10 & 20 & 3 & 0.0654 \\
54 & 10 & 10 & 2 & 0.0655 \\
55 & 10 & 40 & 2 & 0.0656 \\
56 & 10 & 50 & 2 & 0.0657 \\
57 & 30 & 50 & 0 & 0.0657 \\
58 & 20 & 20 & 0 & 0.0657 \\
59 & 20 & 40 & 4 & 0.0659 \\
60 & 10 & 10 & 4 & 0.0661 \\
\bottomrule
\end{tabular}
\end{minipage}

\end{table}


To verify the adequacy of the linearization points used in the linear approximation~\eqref{eq:llp-lin}, we select the top three performing parameter combinations from each buffer size case. The selected sets are $\{(0, 10, 0),\ (10, 10, 0),\ (0, 10, 2)\}$ for buffer size 0 and $\{(0, 10, 0),\ (0, 20, 0),\ (10, 10, 0)\}$ for buffer size 2. 
To include a case where all parameters are nonzero, we add the triplet $(20, 30, 2)$. We compare two settings for the number of linearization points: $(100, 100)$ and $(150, 150)$, where each pair corresponds to the number of points used for $(\DisaggregatedArrivalRate,\ \ArrivalRate)$ respectively. In addition, we include the exact method iTAPAS \citep{XIE2016406}\footnote{Code adapted from \url{https://github.com/hanqiu92/itapas}}, a gradient projection-based algorithm designed for traffic assignment problems, to assess the validity of the linear approximation by evaluating the optimality gap with iTAPAS and the results obtained from each linearization point setting. To handle positive and negative objective values, the optimality gap is calculated as $\text{Gap} = \frac{|\text{Obj}_{\eqref{eq:llp-lin}} - \text{Obj}_{\text{iTAPAS}}|}{\max\left(|\text{Obj}_{\eqref{eq:llp-lin}}|, |\text{Obj}_{\text{iTAPAS}}|, \epsilon\right)}$, where $\epsilon$ is a small tolerance to avoid division by zero, and $\text{Obj}_{\text{method}}$ is the optimal leader's objective value returned by $\text{method}$. The results are presented in Table~\ref{tab:la-accuracy-b=0} and Table~\ref{tab:la-accuracy-b=2}. For both buffer size cases, the linearization point setting (100,100) achieves small optimality gap ($\leq 0.02$) for triplets with $\theta^{-1} = 0$, while the linearization point setting $(150, 150)$ only improves the gap slightly. For triplets with $\theta^{-1} >0$, both linearization point settings achieve larger gaps ($\geq 0.1$). Notably, in these cases, iTAPAS requires over an hour to converge to the user equilibrium solution with a relative gap (as defined in \cite{XIE2016406}) of at most $0.02$, while the linear approximation reaches a solution in under two minutes.

Because of the poor accuracy of \eqref{eq:llp-lin} when $\theta^{-1} >0$, we analyze whether this affects the ranking of disutility parameter triplets under the KL divergence metric—specifically, whether triplets with $\theta^{-1} > 0$ are systematically underestimated. To this end, due to limited computational resources and long running times, we compute the KL divergence using iTAPAS only for the selected combined set of triplets, rather than the full grid search set, as shown in Table~\ref{tab:kl-itapas}. We observe that among the triplets in the combined set, $(0,10,0)$ is still the best one with the lowest KL divergence score. In summary, with both buffer sizes (zero and two), the linearization point setting of $(100, 100)$ is sufficiently accurate for the optimal parameter combinations with $\theta^{-1} =0$, e.g. $(0,10,0)$ and $(10,10,0)$. Although the linear approximation performs poorly when $\theta^{-1} > 0$, we omit these cases from further analysis, as they do not yield the lowest KL divergence scores, as confirmed by both the iTAPAS and linear approximation results. Furthermore, in the remaining experiments, we consider linearization points $(100,100)$ as they offer a good balance between accuracy and solving times.

\begin{table}[h]
\centering
\caption{Linear approximation accuracy | $B = 0$}
\label{tab:la-accuracy-b=0}
\resizebox{\linewidth}{!}{%
\begin{tabular}{c|rrr|rrr|rrr}
\hline
\textbf{Triplet} & \multicolumn{3}{c|}{\textbf{iTAPAS}} & \multicolumn{3}{c|}{\textbf{\ref{eq:llp-lin} (100, 100)}} & \multicolumn{3}{c}{\textbf{\ref{eq:llp-lin} (150, 150)}} \\
$(\alpha,\beta, \theta^{-1})$ & \textbf{Obj. Val} & \textbf{Rel. Gap} & \textbf{Time (s)} & \textbf{Obj. Val} & \textbf{Opt. Gap} & \textbf{Time (s)} & \textbf{Obj. Val} & \textbf{Opt. Gap} & \textbf{Time (s)} \\
\hline
(0, 10, 0) & 526.03 & 0.00 & 82.20 & 518.69 & 0.01 & 42.81 & 522.62 & 0.01 & 63.46 \\
(0, 20, 0) & 806.66 & 0.00 & 109.44 & 790.69 & 0.02 & 42.82 & 800.55 & 0.01 & 65.05 \\
(10, 10, 0) & 868.30 & 0.00 & 69.20 & 860.18 & 0.01 & 42.94 & 865.14 & 0.00 & 63.40 \\
(0, 10, 2) & -479.08 & 0.01 & 3668.69 & -693.47 & 0.31 & 52.69 & -589.11 & 0.19 & 80.84 \\
(20, 30, 2) & 752.14 & 0.01 & 3664.07 & 526.92 & 0.30 & 55.80 & 638.82 & 0.15 & 83.69 \\
\hline
\end{tabular}%
}
\end{table}

\begin{table}[h]
\centering
\caption{Linear approximation accuracy | $B = 2$}
\label{tab:la-accuracy-b=2}
\resizebox{\linewidth}{!}{%
\begin{tabular}{c|rrr|rrr|rrr}
\hline
\textbf{Triplet} & \multicolumn{3}{c|}{\textbf{iTAPAS}} & \multicolumn{3}{c|}{\textbf{\ref{eq:llp-lin} (100, 100)}} & \multicolumn{3}{c}{\textbf{\ref{eq:llp-lin} (150, 150)}} \\
$(\alpha,\beta, \theta^{-1})$ & \textbf{Obj. Val} & \textbf{Rel. Gap} & \textbf{Time (s)} & \textbf{Obj. Val} & \textbf{Opt. Gap} & \textbf{Time (s)} & \textbf{Obj. Val} & \textbf{Opt. Gap} & \textbf{Time (s)} \\
\hline
(0, 10, 0) & 277.50 & 0.00 & 49.40 & 274.15 & 0.01 & 60.10 & 275.80 & 0.01 & 87.28 \\
(0, 20, 0) & 318.38 & 0.00 & 81.42 & 312.06 & 0.02 & 60.09 & 315.56 & 0.01 & 87.04 \\
(10, 10, 0) & 639.75 & 0.00 & 55.73 & 635.67 & 0.01 & 60.02 & 638.05 & 0.00 & 87.27 \\
(0, 10, 2) & -720.91 & 0.02 & 3620.64 & -933.67 & 0.23 & 68.25 & -830.34 & 0.13 & 103.07 \\
(20, 30, 2) & 67.36 & 0.01 & 3643.33 & -158.35 & 1.43 & 76.05 & -46.81 & 1.69 & 115.01 \\
\hline
\end{tabular}%
}
\end{table}

\begin{table}[h]
\centering
\caption{KL divergence comparison with iTAPAS}
\label{tab:kl-itapas}
\begin{tabular}{c|cc|cc}
\toprule
\textbf{Triplet} & \multicolumn{2}{c|}{\textbf{Buffer = 0}} & \multicolumn{2}{c}{\textbf{Buffer = 2}} \\
 $(\alpha,\beta, \theta^{-1})$ & iTAPAS & \ref{eq:llp-lin} (100, 100) & iTAPAS & \ref{eq:llp-lin} (100, 100) \\
\midrule
(0, 10, 0) & 0.0501 & 0.0502 & 0.0313 & 0.0327 \\
(0, 20, 0) & 0.0554 & 0.0702 & 0.0395 & 0.0437 \\
(10, 10, 0) & 0.0522 & 0.0536 & 0.0527 & 0.0498 \\
(0, 10, 2) & 0.0605 & 0.0568 & 0.0610 & 0.0618 \\
(20, 30, 2) & 0.0613 & 0.0612 & 0.0675 & 0.0702 \\
\bottomrule
\end{tabular}
\end{table}

\subsection{Upper-level problem}\label{subsec:casestudy_upperlevel}

\subsubsection{Instance generation}
\label{subsec:upper_instance}
In the previous section, we focused on the modeling of user behaviors assuming the locations of the facilities are fixed. In this section, we showcase the ability of our FLP formulations (Section~\ref{sec:bilevel_formulation_of_facility_location_problem}) to optimize the locations of the facilities taking into account the user behavior.
In particular, we consider \textbf{What-if scenarios} in which existing sites are hypothetically relocated to new locations. This is not meant to suggest that relocation is practically feasible—rather, it serves as an analytical tool to assess potential improvements in our accessibility metric. Given the current facility locations and an integer $X$, the $X$ worst-performing (i.e., with the lowest numbers of charging sessions in the last 30 days) leader sites are allowed to be relocated to new sites. The set of candidate locations, $J \setminus J_1$, includes the population centroids (recall Figure~\ref{fig:original_map_comp}) as well as the $X$ original locations of the stations selected for relocation.

We use the same demand data as we used to study the lower-level problem in Section~\ref{subsec:casestudy_lowerlevel}, with 34 leader sites and 2 competitor sites.
To observe the effect of the competition, we add 7 additional competitor sites at the highest-demand population nodes, which are the 7 triangles overlapping the green demand nodes in Figure~\ref{fig:original_map_comp}. Since the majority of the network consists of L2 2-outlet sites (recall Section~\ref{subsec:casestudy_lowerlevel}), these 7 additional competitor sites and the $X$ leader sites to be relocated are all assumed to be L2 sites with 2 outlets.

Three relocation scenarios were tested: $X = 0$ (baseline, no relocation), $X = 3$ and $X = 6$. 
We also use two different buffer sizes: zero and ten. The former case reflects the current operation, while the latter case is a fictitious scenario to observe the impact of waiting space—an increasingly relevant factor with the emergence of virtual queueing technologies \citep{bellan2025tesla}. 
Two combinations of user disutility parameters $(\alpha, \beta, \theta^{-1})$ were used: $(0, 10, 0)$ and $(10, 10, 0)$, which are the two best triplets for buffer size of zero from the calibration presented in the previous section. The experiments are conducted under two scenarios: without competition and with competition.

\subsubsection{Results} 
In this section, we apply our linearization approach and surrogate-based heuristic to generate high-quality location decisions to problem~\eqref{eq:flp} within the context of the case study scenarios. We implement these methodologies in Python 3.10.16 and call CPLEX 22.1.2 through the docplex package to solve the corresponding optimization models, with a time limit of 3,600 seconds. When this limit is reached, we use the best feasible (incumbent) solution found by CPLEX up to that point, and we report the optimality gap given by this solver. All computations were performed on an Intel(R) Xeon(R) Gold 6226 CPU clocked at 2.70 GHz, with CPLEX restricted to 16 threads.

These experiments yield several key insights regarding the impact of relocation, buffer capacity, competition, and model formulation on system performance. Performance is evaluated using the leader's Total True Throughput Rate (TTR) and the Average True Throughput Rate (ATR). The ATR, calculated as the total true throughput divided by the number of leader sites (i.e., 34), provides a more intuitive measure of throughput rate per site. The term ``true'' refers to throughput values $\EffectiveArrivalRate$, in contrast to the approximated throughput used in the objective function~\eqref{eq:linearisation_approach_with_linear_objective}. The results comparing the solving of our linearization approach (Section~\ref{subsec:linearization_approach}) with the heuristic (Section~\ref{subsec:surrogate_based_heuristic}) are presented in Table~\ref{tab:wis-no-comp} and Table~\ref{tab:wis-with-comp} for the case without competition and with competition, respectively. We should recall that the linearization approach solution is driven by the objective of maximizing throughput, while that of the heuristic is driven by the lower-level objective; thus, their optimal objective values (not included in the tables) are not comparable.

\begin{table}[h]
\centering
\caption{What-if scenarios (without competition)}
\label{tab:wis-no-comp}
\resizebox{\linewidth}{!}{%
\begin{tabular}{rrrrrrrrr}
\toprule
Relocation & Buffer & Triplet & Model & Opt. Gap & TTR & ATR & Elapsed Time & Solving Time \\
\midrule
0 & 0 & (0, 10, 0) & Linearization & 0.00 & 251.10 & 7.39 & 265.44 & 11.66 \\
0 & 0 & (0, 10, 0) & Heuristic     & 0.00 & 251.10 & 7.39 & 201.87 & 1.93 \\
0 & 0 & (10, 10, 0) & Linearization & 0.00 & 251.30 & 7.39 & 264.20 & 13.03 \\
0 & 0 & (10, 10, 0) & Heuristic     & 0.00 & 251.30 & 7.39 & 198.13 & 1.94 \\
0 & 10 & (0, 10, 0) & Linearization & 0.00 & 318.72 & 9.37 & 465.86 & 11.64 \\
0 & 10 & (0, 10, 0) & Heuristic     & 0.00 & 318.72 & 9.37 & 404.39 & 1.86 \\
0 & 10 & (10, 10, 0) & Linearization & 0.00 & 320.92 & 9.44 & 486.29 & 13.55 \\
0 & 10 & (10, 10, 0) & Heuristic     & 0.00 & 320.92 & 9.44 & 389.47 & 1.90 \\
3 & 0 & (0, 10, 0) & Linearization & 0.01 & 251.71 & 7.40 & 3856.10 & 3601.36 \\
3 & 0 & (0, 10, 0) & Heuristic     & 0.00 & 251.53 & 7.40 & 230.63 & 39.50 \\
3 & 0 & (10, 10, 0) & Linearization & 0.01 & 252.00 & 7.41 & 3857.78 & 3604.44 \\
3 & 0 & (10, 10, 0) & Heuristic     & 0.00 & 251.79 & 7.41 & 225.21 & 31.72 \\
3 & 10 & (0, 10, 0) & Linearization & 0.00 & 320.72 & 9.43 & 4063.85 & 3602.74 \\
3 & 10 & (0, 10, 0) & Heuristic     & 0.00 & 319.90 & 9.41 & 401.25 & 3.74 \\
3 & 10 & (10, 10, 0) & Linearization & 0.00 & 321.25 & 9.45 & 4066.16 & 3601.01 \\
3 & 10 & (10, 10, 0) & Heuristic     & 0.00 & 321.14 & 9.45 & 423.68 & 23.52 \\
6 & 0 & (0, 10, 0) & Linearization & 0.01 & 251.84 & 7.41 & 3854.48 & 3600.94 \\
6 & 0 & (0, 10, 0) & Heuristic     & 0.00 & 251.53 & 7.40 & 241.72 & 46.76 \\
6 & 0 & (10, 10, 0) & Linearization & 0.01 & 252.14 & 7.42 & 3863.08 & 3601.02 \\
6 & 0 & (10, 10, 0) & Heuristic     & 0.00 & 251.80 & 7.41 & 249.38 & 48.26 \\
6 & 10 & (0, 10, 0) & Linearization & 0.00 & 320.98 & 9.44 & 4096.32 & 3611.09 \\
6 & 10 & (0, 10, 0) & Heuristic     & 0.00 & 319.68 & 9.40 & 404.07 & 3.10 \\
6 & 10 & (10, 10, 0) & Linearization & 0.00 & 321.29 & 9.45 & 4085.05 & 3618.27 \\
6 & 10 & (10, 10, 0) & Heuristic     & 0.00 & 321.18 & 9.45 & 431.81 & 30.36 \\
\bottomrule
\end{tabular}%
}
\end{table}

\begin{table}[h]
\centering
\caption{What-if scenarios (with competition)}
\label{tab:wis-with-comp}
\resizebox{\linewidth}{!}{%
\begin{tabular}{rrrrrrrrr}
\toprule
Relocation & Buffer & Triplet & Model & Opt. Gap & TTR & ATR & Elapsed Time & Solving Time \\
\midrule
0 & 0 & (0, 10, 0) & Linearization & 0.00 & 217.03 & 6.38 & 275.65 & 11.85 \\
0 & 0 & (0, 10, 0) & Heuristic     & 0.00 & 217.03 & 6.38 & 204.94 & 2.15 \\
0 & 0 & (10, 10, 0) & Linearization & 0.00 & 217.38 & 6.39 & 272.16 & 14.13 \\
0 & 0 & (10, 10, 0) & Heuristic     & 0.00 & 217.38 & 6.39 & 206.48 & 2.22 \\
0 & 10 & (0, 10, 0) & Linearization & 0.00 & 270.17 & 7.95 & 491.35 & 12.32 \\
0 & 10 & (0, 10, 0) & Heuristic     & 0.00 & 270.17 & 7.95 & 410.20 & 2.46 \\
0 & 10 & (10, 10, 0) & Linearization & 0.00 & 271.22 & 7.98 & 483.48 & 14.23 \\
0 & 10 & (10, 10, 0) & Heuristic     & 0.00 & 271.21 & 7.98 & 409.62 & 2.18 \\
3 & 0 & (0, 10, 0) & Linearization & 0.09 & 220.67 & 6.49 & 3869.98 & 3602.54 \\
3 & 0 & (0, 10, 0) & Heuristic     & 0.00 & 217.74 & 6.40 & 256.00 & 49.48 \\
3 & 0 & (10, 10, 0) & Linearization & 0.08 & 222.17 & 6.53 & 3871.88 & 3606.12 \\
3 & 0 & (10, 10, 0) & Heuristic     & 0.00 & 218.14 & 6.42 & 250.34 & 44.86 \\
3 & 10 & (0, 10, 0) & Linearization & 0.06 & 284.64 & 8.37 & 4146.78 & 3605.29 \\
3 & 10 & (0, 10, 0) & Heuristic     & 0.00 & 266.78 & 7.85 & 411.78 & 2.88 \\
3 & 10 & (10, 10, 0) & Linearization & 0.10 & 282.82 & 8.32 & 4101.59 & 3604.21 \\
3 & 10 & (10, 10, 0) & Heuristic     & 0.00 & 272.23 & 8.01 & 441.91 & 22.41 \\
6 & 0 & (0, 10, 0) & Linearization & 0.14 & 221.51 & 6.51 & 3864.37 & 3601.31 \\
6 & 0 & (0, 10, 0) & Heuristic     & 0.00 & 218.03 & 6.41 & 238.19 & 41.47 \\
6 & 0 & (10, 10, 0) & Linearization & 0.13 & 223.06 & 6.56 & 3878.59 & 3604.59 \\
6 & 0 & (10, 10, 0) & Heuristic     & 0.00 & 218.71 & 6.43 & 242.13 & 42.11 \\
6 & 10 & (0, 10, 0) & Linearization & 0.08 & 298.19 & 8.77 & 4081.41 & 3602.58 \\
6 & 10 & (0, 10, 0) & Heuristic     & 0.00 & 264.68 & 7.78 & 418.98 & 3.20 \\
6 & 10 & (10, 10, 0) & Linearization & 0.10 & 292.50 & 8.60 & 4095.09 & 3610.75 \\
6 & 10 & (10, 10, 0) & Heuristic     & 0.00 & 271.64 & 7.99 & 439.28 & 23.22 \\
\bottomrule
\end{tabular}%
}
\end{table}

\paragraph{\textbf{Comparison of computational performance}} 
In comparing computational performance, it is important to distinguish between \emph{solving time} and \emph{elapsed time}: the former refers to the time CPLEX reports for solving the model, while the latter includes all overhead such as model construction and preprocessing.
In terms of elapsed and solving time, the solving of the heuristic model significantly outperforms that of the linearization model. This is expected due to its simpler formulation, namely absence of the lower-level dual formulation, SOS2 and indicator constraints. Additionally, when relocation is allowed, the linearization model consistently reaches the imposed time limit. In the absence of competition, it typically achieves near-zero optimality gaps (i.e., $\leq 0.01$). However, under competition, the optimality gap is notably larger (up to $\leq 0.14$), likely due to the increased complexity introduced by a larger lower-level that now considers also the competitor locations.
 
\paragraph{\textbf{Comparison of TTR}}
When no relocation is allowed ($X = 0$), both models perform identically in terms of TTR (and ATR) because they reduce to the lower-level problem that simply optimizes user flows over fixed site locations. However, when relocation is enabled ($X > 0$), the linearization model consistently achieves higher TTR than the surrogate-based heuristic solution, with the advantage being significative under competition (e.g. when $X=6$, $B=10$ and $(\alpha, \beta, \theta^{-1})=(0,10,0)$ the linearization solution is 11\% better). On one hand, in the absence of competition, both models achieve solutions with similar TTR because their objectives partially align; that is, maximizing throughput generally benefits users. On the other hand, under competition, the surrogate-based heuristic model prioritizes user utility, which explicitly accounts for the competitor's locations, whereas the linearization model focuses on maximizing the TTR over the leader's locations. The TTR gap between both models also widens with increased buffer size, as larger buffers amplify the divergence in their objectives. Notably, the heuristic model does not improve TTR monotonically with additional relocation. For example, under competition, with parameter combination $(0, 10, 0)$ and a buffer size of ten, the relocation scenario $X = 6$ results in lower TTR compared to both $X = 0$ and $X = 3$. As a result, the remainder of this analysis focuses exclusively on the solutions of the linearization model. That said, the heuristic approach may be preferred in practice for its faster runtime. A network solution analysis of both models is also included later to provide additional insight.

\paragraph{\textbf{Impact of user parameters}}
Across all experiments, the solutions associated with the parameter combination $(10, 10, 0)$ generally lead to slightly better TTR than the ones associated with $(0, 10, 0)$, with a typical difference of no more than 2 users/day in TTR. An exception arises under competition, particularly when the buffer size is ten and relocation is allowed at $X = 3$ and $X = 6$, where the solution for $(0, 10, 0)$ outperforms the one of $(10, 10, 0)$ by approximately 2 and 6 users/day, respectively. This can be intuitively explained by the zero weight on average waiting time ($\alpha = 0$), which reflects an indifference to waiting. In such cases, increasing the buffer size has a greater impact on throughput, since users care more about the availability of space in the queue than about the time spent waiting. However, we should be cautious with these observations as we are not necessarily analyzing the optimal solution. The existence of cases with non-zero optimality gap, reflects that the solution under analyzes was not proven to be optimal. Therefore, we refrain from drawing definitive conclusions about the impact of one parameter setting over the other. The subsequent analysis applies to both parameter combinations, and we focus on $(0, 10, 0)$ in the rest of this section.

\paragraph{\textbf{Impact of relocation budget}}
Our results demonstrate that relocating sites results in a clear improvement in TTR compared to the original network configuration with no relocation, even when the linearization method attains the time limit of 1h and optimality gap is as high as $0.08$ or more. Increasing the number of relocated sites from $X = 3$ to $X = 6$ generally leads to further gains in TTR. However, the marginal improvement is smaller than the initial jump from $X = 0$ to $X = 3$, indicating potential diminishing returns. 


\paragraph{\textbf{Synergy of relocation and buffer size increase}}  
With competition and the parameter combination $(0, 10, 0)$, increasing the buffer size alone (with no relocation) improves ATR by approximately 1.57 users/day, while allowing relocation of $X = 3$ alone (with buffer size zero) yields a more modest ATR improvement of about 0.11 users/day. However, when both strategies are combined—buffer size of ten and relocation of $X = 3$—the ATR increases by approximately 1.99 users/day, exceeding the sum of their individual effects. This suggests a synergistic interaction between buffer capacity and relocation when using the location decisions of the linearization model under competition, where buffer size amplifies the effectiveness of relocation. A similar interaction was observed earlier in TTR performance, where the linearization model outperformed the heuristic model more significantly with larger buffer sizes—further reinforcing the role of buffer capacity as a performance multiplier under competitive dynamics. Without competition, this synergistic effect may still be present but is less pronounced, and is more evident in TTR rather than in ATR, due to the small differences. One possible explanation is that, in the absence of competition and with a relatively high demand factor of $0.3\%$, users have no choice but to use the available sites, regardless of relocation. 

\begin{figure}[h!]
    \centering
    \includegraphics[width=\textwidth]{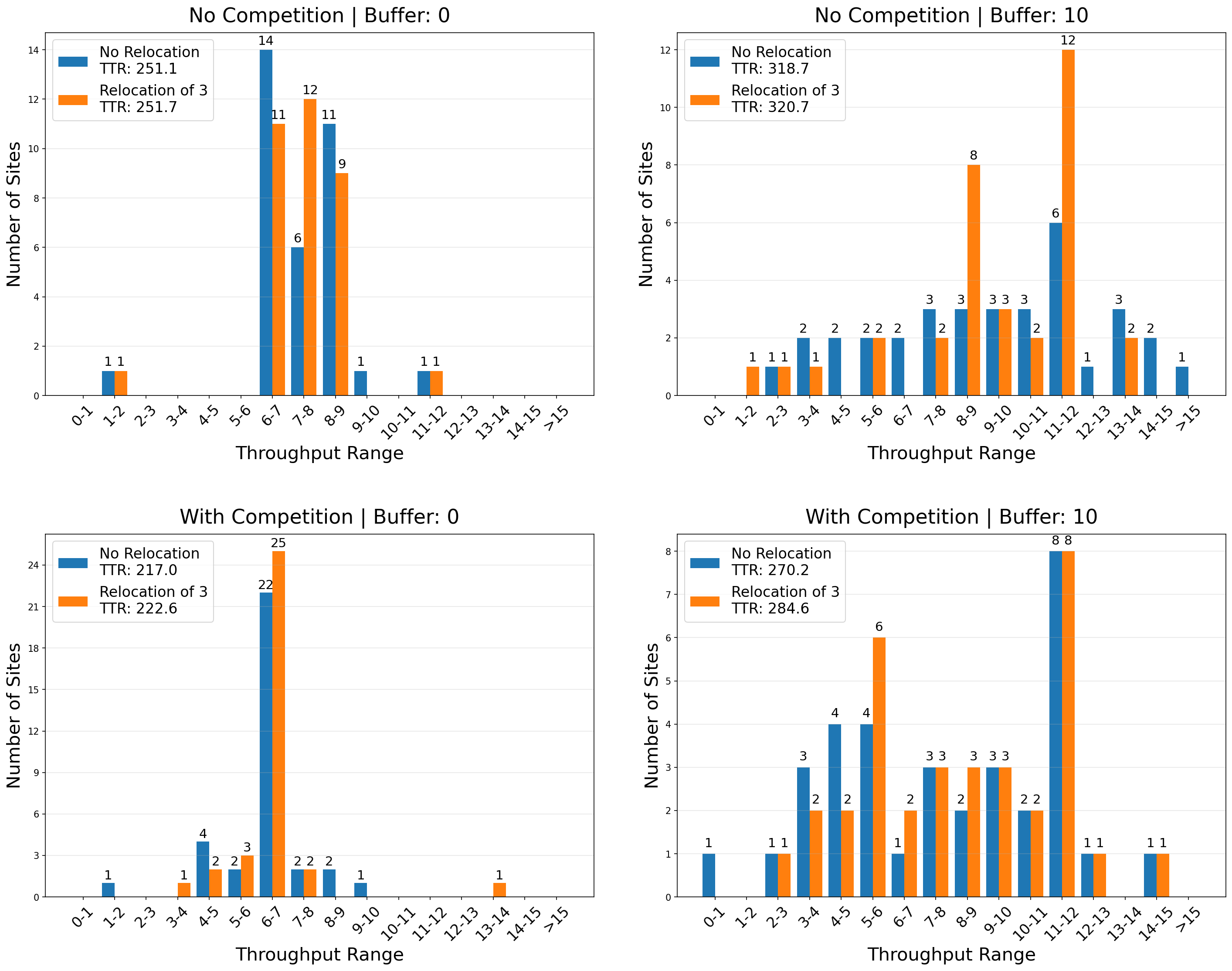}
    \caption{Distribution of throughput across sites under various settings: relocation of $X \in \{0, 3\}$, $B \in \{0, 10\}$, without or with competition, and parameter triplet $(0, 10, 0)$.}
    \label{fig:throughput_barcharts_reloc3}
\end{figure}

\paragraph{\textbf{Throughput distribution}} Figure~\ref{fig:throughput_barcharts_reloc3} compares the distribution of true throughput rates across all sites for the parameter combination $(0, 10, 0)$ under relocation scenarios $X = 0$ and $X = 3$, using the location decisions of the linearization model. Beyond the overall TTR trends previously discussed, the figure reveals that increasing the buffer size to ten shifts the throughput distribution toward higher values—reflecting the ability to handle more users due to additional waiting space. Furthermore, in the absence of competition and both buffer size settings, relocation tends to shift more sites toward the higher end of the throughput distribution compared to the no-relocation scenario. In contrast, under competition, relocation increases the number of sites operating at moderate throughput levels, likely because users are more spread out across the network when they have more options to choose from.

\begin{figure}[h!]
\centering

\begin{minipage}[t]{0.48\textwidth}
    \centering
    \includegraphics[width=\linewidth, height=0.4\textheight, keepaspectratio]{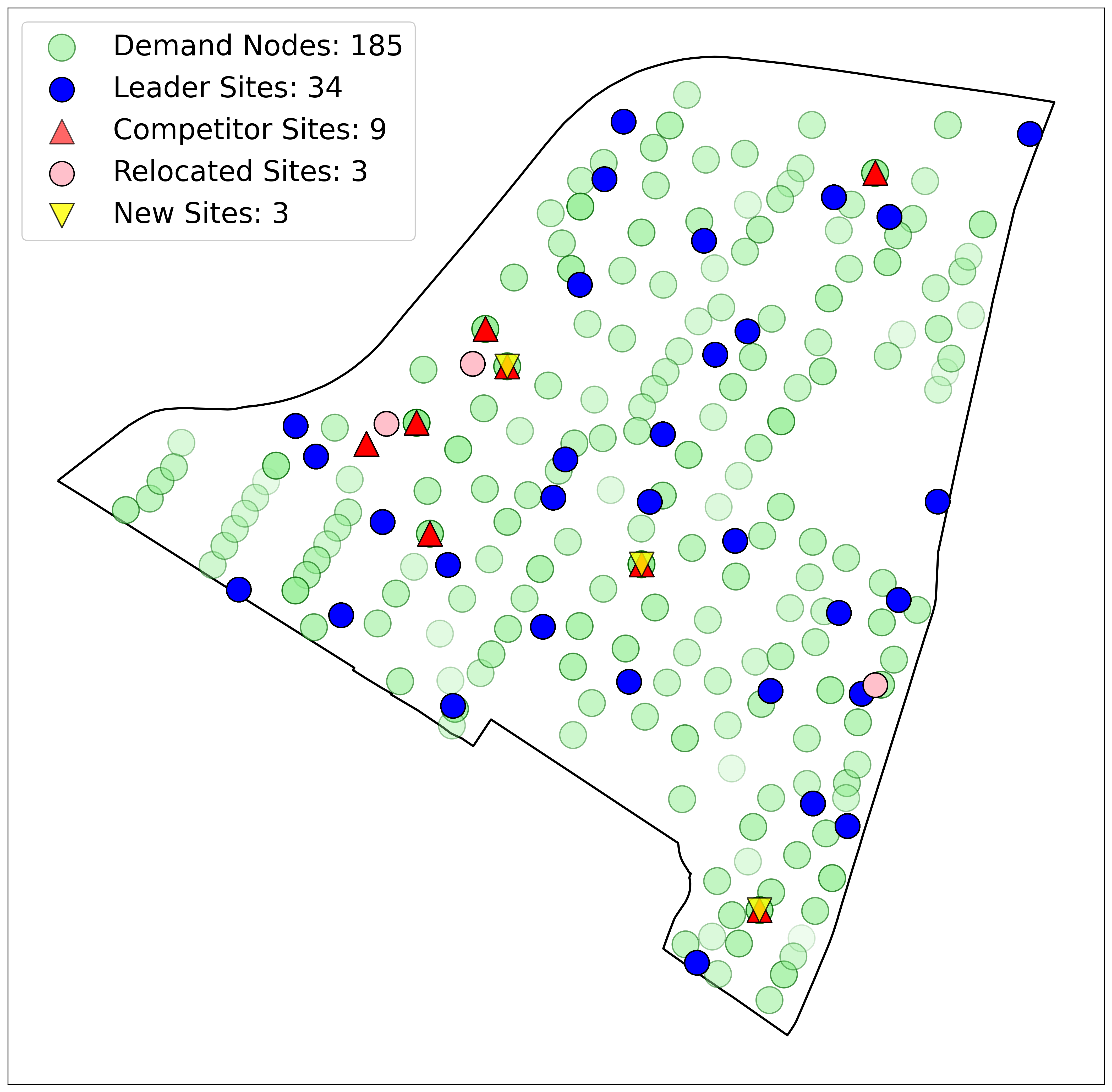}
    \caption{Linearization and $(0, 10, 0)$}
    \label{fig:throughput_a0}
\end{minipage}\hfill
\begin{minipage}[t]{0.49\textwidth}
    \centering
    \includegraphics[width=\linewidth, height=0.4\textheight, keepaspectratio]{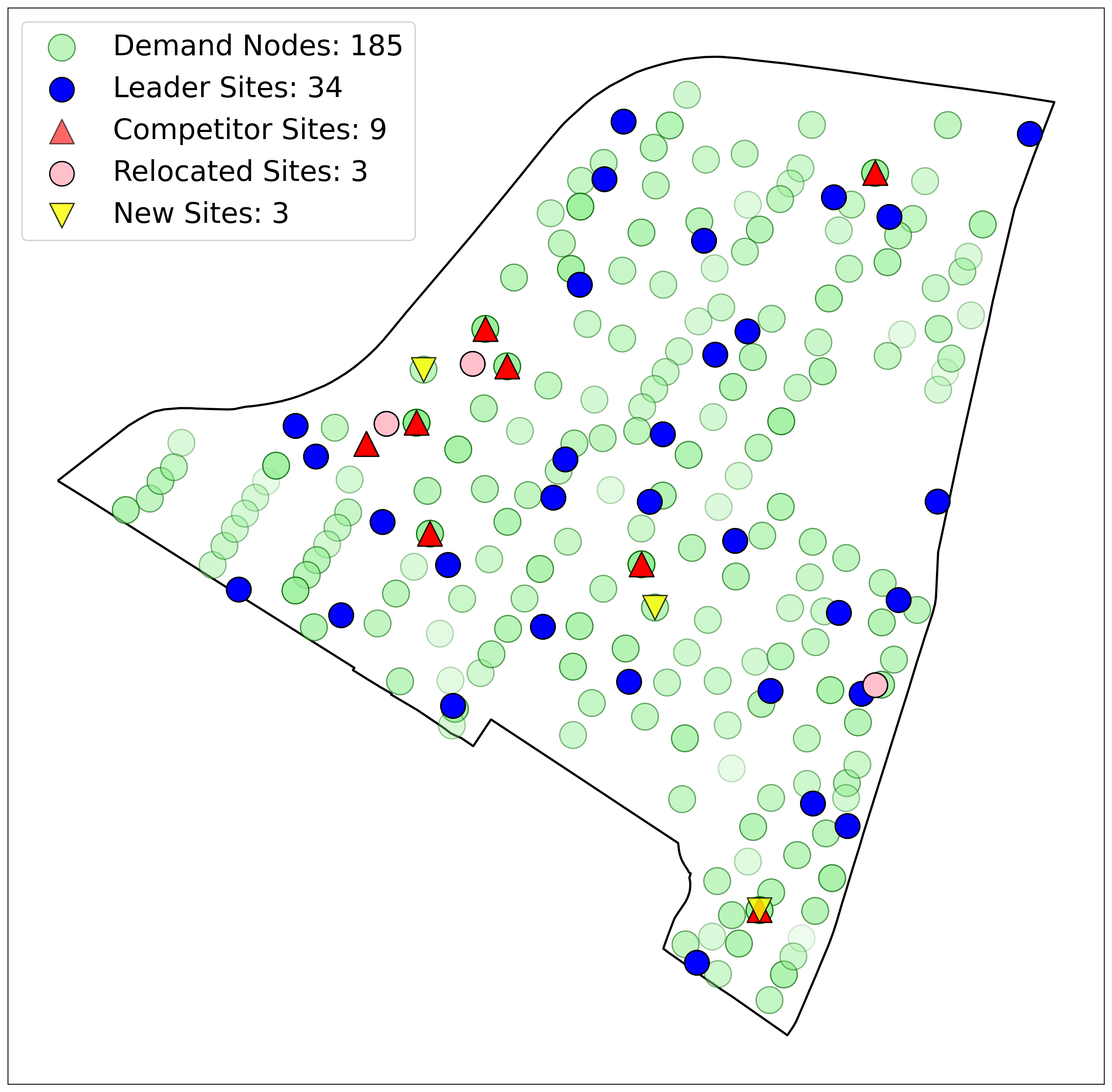}
    \caption{Linearization and $(10, 10, 0)$}
    \label{fig:throughput_a10}
\end{minipage}

\vspace{0.2cm}

\begin{minipage}[t]{0.48\textwidth}
    \centering
    \includegraphics[width=\linewidth, height=0.4\textheight, keepaspectratio]{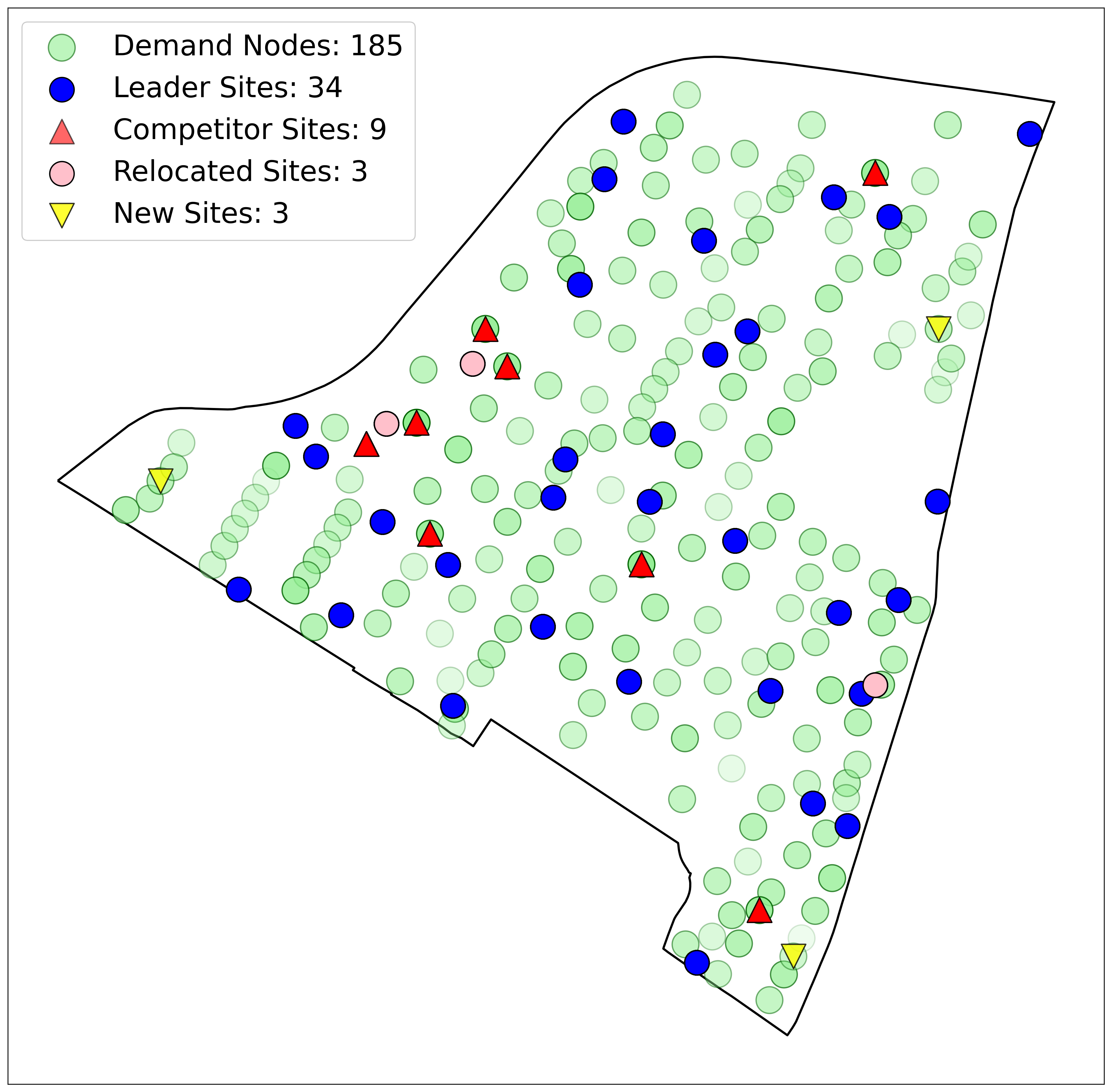}
    \caption{Heuristic and $(0, 10, 0)$}
    \label{fig:disutility_a0}
\end{minipage}\hfill
\begin{minipage}[t]{0.49\textwidth}
    \centering
    \includegraphics[width=\linewidth, height=0.4\textheight, keepaspectratio]{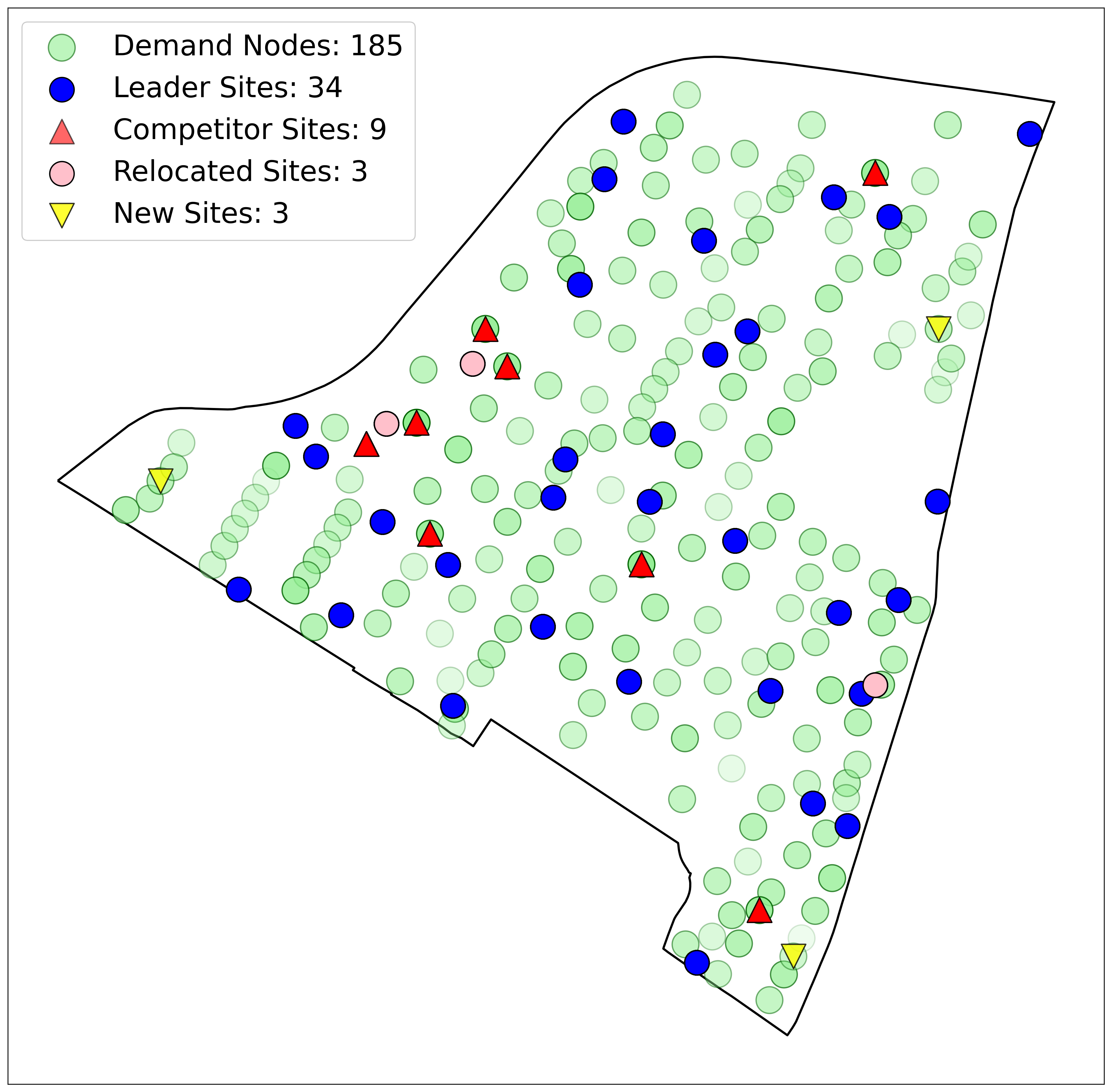}
    \caption{Heuristic and $(10, 10, 0)$}
    \label{fig:disutility_a10}
\end{minipage}

\caption{Networks comparison of linearization and heuristic models under competition, for relocation parameter $X=3$ and buffer size $B=0$. Triplets $(\alpha, \beta, \theta^{-1})$ denote disutility parameters.}
\label{fig:maps_comparison}
\end{figure}

\paragraph{\textbf{Network analysis}} 
Building on our quantitative results, we now examine the resulting networks across different scenarios to better understand the behavior and implications of the models.
Figure~\ref{fig:maps_comparison} displays four networks under competition associated with the location solutions computed for the linearization and heuristic models, with relocation parameter $X=3$, no buffer capacity ($B=0$) and two user parameter combinations: $(0, 10, 0)$ and $(10, 10, 0)$. The three pink circles indicate existing leader sites that are relocated, while the three yellow inverted triangles represent new sites added by our models.

For the linearization model, (Figures~\ref{fig:throughput_a0} and~\ref{fig:throughput_a10}), the selected sites are concentrated around heavily congested regions. This aligns with the model's objective of maximizing throughput, as it tends to favor locations with high demand. We recall from Section~\ref{subsec:upper_instance} that 7 additional competitor sites are placed at the highest-demand population nodes, visually represented as red triangles overlapping with green demand nodes. Notably, the solution under the triplet $(0, 10, 0)$ (Figure~\ref{fig:throughput_a0}) places new sites directly on 3 of these 7 additional competitor locations. In contrast, the triplet $(10, 10, 0)$ setting (Figure~\ref{fig:throughput_a10}) results in new sites that remain within the same congested area but are not necessarily located at the highest-demand nodes. This may be explained by the nonzero $\alpha$ parameter in $(10, 10, 0)$, which captures users’ tolerance for waiting. As a result, travel time becomes a relatively smaller component of the user disutility, allowing optimal sites to be placed near—but not necessarily on—high-demand nodes, as long as waiting time and balking remain relatively low.

The solutions for the heuristic model show identical new site locations that are spread out across the region for both parameter triplets (Figures~\ref{fig:disutility_a0} and \ref{fig:disutility_a10}). This spreading suggests that the heuristic prioritizes placing stations to cover as many demand nodes as possible in order to serve the full user base, which in this case results in broad spatial coverage. The observed invariance may simply be a coincidence due to the small number of relocated sites ($X=3$).

Finally, the decisions resulting from the linearization and heuristic models appear fundamentally different, further underscoring the significant impact of the objective functions driving each approach.

\begin{table}[h]
\centering
\caption{Robustness test of linearization model under competition, for relocation parameter $X=3$ and buffer size $B=0$. Triplets $(\alpha, \beta, \theta^{-1})$ denote disutility parameters.}
\label{tab:robust_X=3}
\resizebox{\textwidth}{!}{%
\begin{tabular}{l r | rrr rrr}
\midrule
\textbf{Disutility Setting} & \textbf{TTR} &
\multicolumn{6}{c}{\textbf{Optimality Gap}} \\
\cmidrule(lr){3-8}
& &
\multicolumn{3}{c}{\textbf{$M/M/1/1+B$}} &
\multicolumn{3}{c}{\textbf{$M/M/s/s+B$}} \\
\cmidrule(lr){3-5} \cmidrule(lr){6-8}
& & (0,0,0) & (0,10,0) & (10,10,0) & (0,0,0) & (0,10,0) & (10,10,0) \\
\midrule
$M/M/s/s+B$ + (0,0,0)   & 223.65 & 0.00\% & 0.60\% & 2.57\% & 0.00\% & 2.03\% & 2.49\% \\
$M/M/s/s+B$ + (0,10,0)  & 220.67 & 0.80\% & 0.22\% & 0.92\% & 0.80\% & 0.00\% & 0.23\% \\
$M/M/s/s+B$ + (10,10,0) & 222.17 & 0.56\% & 0.00\% & 0.53\% & 0.56\% & 0.61\% & 0.00\% \\
\midrule
\end{tabular}
} 
\end{table}

\begin{table}[h]
\centering
\caption{Robustness test of the linearization model under competition, for relocation parameter $X=3$ and buffer size $B=0$. Triplets $(\alpha, \beta, \theta^{-1})$ denote disutility parameters. The number of servers $s$ is set to four for all sites.}
\label{tab:robust_X=3_s=4}
\resizebox{\textwidth}{!}{%
\begin{tabular}{l r | rrr rrr}
\midrule
\textbf{Disutility Setting} & \textbf{TTR} &
\multicolumn{6}{c}{\textbf{Optimality Gap}} \\
\cmidrule(lr){3-8}
& &
\multicolumn{3}{c}{\textbf{$M/M/1/1+B$}} &
\multicolumn{3}{c}{\textbf{$M/M/s/s+B$}} \\
\cmidrule(lr){3-5} \cmidrule(lr){6-8}
& & (0,0,0) & (0,10,0) & (10,10,0) & (0,0,0) & (0,10,0) & (10,10,0) \\
\midrule
$M/M/s/s+B$ + (0,0,0)   & 280.59 & 0.00\% & 0.75\% & 2.56\% & 0.00\% & 0.13\% & 0.13\% \\
$M/M/s/s+B$ + (0,10,0)  & 274.58 & 0.39\% & 0.24\% & 1.49\% & 0.39\% & 0.00\% & 0.00\% \\
$M/M/s/s+B$ + (10,10,0) & 283.73 & 1.47\% & 0.23\% & 1.44\% & 1.47\% & 0.00\% & 0.00\% \\
\midrule
\end{tabular}
} 
\end{table}

\paragraph{\textbf{Robustness test}}

To assess the impact of disutility settings on the optimal decisions and TTR of the linearization model, we conduct a robustness test. Let $\mathcal{A}$ denote the set of disutility settings, each defined by a combination of a queueing system (either $M/M/1/1+B$ or $M/M/s/s+B, s \ge 2$) (recall the definitions from Section~\ref{subsec:queues_comparison}) and a user parameters combination in $\{ (0, 0, 0), (0, 10, 0), (10, 10, 0)\}$, where $(0, 0, 0)$ indicates that users selects the closest stations. For each setting $a \in \mathcal{A}$, we compute the optimal solution $\Build_a$, obtained by maximizing the objective function $\EffectiveArrivalRateFunction_a(\Build)$. Then, for each pair $(a, a') \in \mathcal{A} \times \mathcal{A}$, we evaluate the achieved TTR of $\Build_a$ under setting $a'$ by calculating $\EffectiveArrivalRateFunction_{a'}(\Build_a)$. This allows us to measure how well each solution performs outside its original disutility setting—that is, how robust the solution is to changes in modeling assumptions. The optimality gap is calculated by
$\text{Optimality Gap} = \dfrac{\EffectiveArrivalRateFunction_{a'}(\Build_{a'}) - \EffectiveArrivalRateFunction_{a'}(\Build_{a})}{\EffectiveArrivalRateFunction_{a'}(\Build_{a'})}$,
where $\Build_{a'}$ is the optimal solution obtained by maximizing the objective function $\EffectiveArrivalRateFunction_{a'}(\Build)$. The comparative results of this robustness test are reported in Table~\ref{tab:robust_X=3} with relocation parameter $X=3$. The rows represent the disutility settings $a'$ with TTR as $\EffectiveArrivalRateFunction_{a'}(\Build_{a'})$, while the columns correspond to the optimality gap of $\EffectiveArrivalRateFunction_{a'}(\Build_{a})$. Since $M/M/s/s+B$ with $s \ge 1$ represents a more realistic queueing model compared to $M/M/1/1+B$, we include only the disutility settings $a'$ evaluated under the $M/M/s/s+B$ setting.

A general observation from Table~\ref{tab:robust_X=3} is that nonzero performance gaps emerge across nearly all settings, indicating noticeable differences in TTR when each solution is applied outside its original disutility setting. In particular, the discrepancies observed between queueing assumptions suggest that simplified models—such as approximating a multi-server system with a single-server queue—should be avoided, as they risk misrepresenting key aspects of system dynamics, namely balking probability and average waiting time. For further analysis, we focus exclusively on solutions from setting $a$ with $M/M/s/s+B$ queue.

The solution from the parameter combination $(0, 0, 0)$ consistently achieves low optimality gaps, while both solutions from $(0, 10, 0)$ and $(10, 10, 0)$ perform poorly when evaluated under the $(0, 0, 0)$ setting. This observation is reasonable, given that $(0, 0, 0)$ is commonly used in the existing literature and serves as a reliable baseline or a “vanilla” assumption. However, if the user parameters are believed to be well-calibrated—based on empirical data from charging sessions—then the solutions from $(0, 10, 0)$ and $(10, 10, 0)$ remain the most appropriate choices within their respective disutility settings $a'$. 

While this case study focuses on the common configuration of two outlets per station, stations with a larger number of outlets can further amplify the performance gap between single-server and multi-server queueing models (see Section~\ref{subsec:queues_comparison}). To investigate this, we conduct an additional robustness test, setting the number of servers per site $s$ to a fictitious value of four. The results are shown in Table~\ref{tab:robust_X=3_s=4}.

As expected, using the $M/M/1/1+B$ model as an approximation of the more realistic $M/M/s/s+B$ queue leads to larger gaps and inferior performance. Furthermore, in the $M/M/s/s+B$ setting, unlike in Table~\ref{tab:robust_X=3}, the triplets $(0,10,0)$ and $(10,10,0)$ yield identical and consistently low optimality gaps, indicating greater robustness compared to $(0,0,0)$.

Consequently, we emphasize the importance of modeling the lower-level problem—namely, user equilibrium—as accurately as possible, due to its direct influence on the leader’s decision-making and the system’s overall throughput.

\section{Conclusion and future work}\label{sec:conclusion}

Our study presents an advanced facility location framework tailored to the operational realities of EV charging in congested and competitive urban environments. We apply a user choice model that integrates queueing performance into a multinomial logit model, allowing users to select charging stations based on travel time, waiting time, and balking probability.

For different queueing systems, where possible, we begin by deriving closed-form expressions for queueing-based metrics, notably the average waiting time and balking probability. Then, our experiments show that the $M/M/s/K$ queue best approximates an empirically validated model for EV charging.

Assuming that users select the station that minimizes their disutility, we show that the user equilibrium problem can be formulated as a convex optimization problem under the assumption that the facilities are modeled as $M/M/s/K$ queues. This property allows for a tractable linear approximation of user behavior, which we incorporate into a bilevel optimization framework where a facility planner (leader) places charging stations in the presence of competing infrastructure. We solve this problem using both a linearized single-level reformulation—aimed at maximizing the throughput—and a surrogate-based heuristic that prioritizes user preferences.

Our results in a real-world scenario demonstrate that the solving of our linearization model leads to (re)locations that outperform the existing network under competition and congestion. Moreover, the increase in throughput resulting from our approach's location decisions becomes particularly significant as buffer size increases, i.e. waiting space expands. This scenario is increasingly plausible due to emerging virtual queueing technologies and is especially relevant for fast-charging stations, where the presence of effective physical queues is more reasonable.

Beyond performance improvements, our findings yield conceptual insights. Simplifying assumptions—such as using single-server approximations for multi-server systems—can significantly distort both user behavior and optimal infrastructure decisions. Furthermore, we find that assumptions regarding user disutility parameters, buffer sizes, and competition levels meaningfully impact the leader’s strategic choices. In practice, this suggests that providers like Circuit Électrique should invest in understanding their competitive environment and the behavioral drivers of user demand. 

This work makes simplifying assumptions that enable analytical tractability but may limit generalizability.  In particular, the use of the multinomial logit model imposes a strong assumption on user choice behavior that may not fully capture real-world complexities. While service rates are known, demand is estimated, and data on queueing behavior—such as balking—is limited or unavailable. We also assume fixed user preferences, travel times, and buffer sizes, despite these factors being unobservable or variable in practice. These assumptions may affect the accuracy of model predictions, particularly in dynamic or heterogeneous environments. Hence, future work should address these limitations by obtaining real-world data on arrivals and waiting times, exploring the scalability of our linear approximation—especially when extending the model to include capacity decisions such as the number of outlets per station—and considering dynamic competition over time.

\section*{Acknowledgments}

The authors sincerely appreciate the support provided by Ribal Atallah of the Hydro-Québec Research Institute for sharing his expertise on electric vehicle charging infrastructure and the associated network. We also thank Steven Lamontagne and İsmail Sevim for their helpful discussions and insightful perspectives related to the project.

This research was supported by Mitacs and Hydro-Québec through the Mitacs Accelerate Program, and partially by the FRQ-IVADO Research Chair in
Data Science for Combinatorial Game Theory.


\bibliographystyle{apalike}
\def\bibname{References}
\bibliography{Ref_paper}

\begin{thebibliography}{}

\bibitem[Aboolian and Karimi, 2025]{ABOOLIAN2025107004}
Aboolian, R. and Karimi, M. (2025).
\newblock Benefit maximizing network design in the public sector.
\newblock {\em Computers \& Operations Research}, 178:107004.

\bibitem[Al-Dahabreh et~al., 2023]{aldahabreh}
Al-Dahabreh, N., Sayed, M.~A., Sarieddine, K., Elhattab, M., Khabbaz, M.~J.,
  Atallah, R.~F., and Assi, C. (2023).
\newblock A data-driven framework for improving public {EV} charging
  infrastructure: Modeling and forecasting.
\newblock {\em IEEE Transactions on Intelligent Transportation Systems}, pages
  1--14.

\bibitem[Anjos et~al., 2020]{ANJOS2020263}
Anjos, M.~F., Gendron, B., and Joyce-Moniz, M. (2020).
\newblock {Increasing electric vehicle adoption through the optimal deployment
  of fast-charging stations for local and long-distance travel}.
\newblock {\em European Journal of Operational Research}, 285(1):263--278.

\bibitem[Arslan and Kara{\c{s}}an, 2016]{arslan2016benders}
Arslan, O. and Kara{\c{s}}an, O.~E. (2016).
\newblock A {B}enders decomposition approach for the charging station location
  problem with plug-in hybrid electric vehicles.
\newblock {\em Transportation Research Part B: Methodological}, 93:670--695.

\bibitem[Beckmann et~al., 1955]{BeckmannEtAl1955}
Beckmann, M.~J., McGuire, C.~B., and Winsten, C.~B. (1955).
\newblock {\em Studies in the Economics of Transportation}.
\newblock RAND Corporation, Santa Monica, CA.

\bibitem[Bellan, 2025]{bellan2025tesla}
Bellan, R. (2025).
\newblock Tesla to test virtual queues at supercharging locations.
\newblock Link:
  \url{https://techcrunch.com/2025/02/20/tesla-to-test-virtual-queues-at-supercharging-locations/}.
  Accessed: 2025-05-26.

\bibitem[Calvo-Jurado et~al., 2024]{CALVOJURADO2024105719}
Calvo-Jurado, C., Ceballos-Martínez, J.~M., García-Merino, J.~C.,
  Muñoz-Solano, M., and Sánchez-Herrera, F.~J. (2024).
\newblock {Optimal location of electric vehicle charging stations using
  proximity diagrams}.
\newblock {\em Sustainable Cities and Society}, 113:105719.

\bibitem[Canada, 2021]{StatisticsCanada2021}
Canada, S. (2021).
\newblock Census profile, 2021 census of population.
\newblock Link:
  \url{https://www12.statcan.gc.ca/census-recensement/2021/dp-pd/prof/about-apropos/about-apropos.cfm?Lang=E}.

\bibitem[Dan and Marcotte, 2019]{dan19}
Dan, T. and Marcotte, P. (2019).
\newblock {Competitive facility location with selfish users and queues}.
\newblock {\em Operations Research}.

\bibitem[D’Ambrosio et~al., 2010]{DAMBROSIO201039}
D’Ambrosio, C., Lodi, A., and Martello, S. (2010).
\newblock {Piecewise linear approximation of functions of two variables in MILP
  models}.
\newblock {\em Operations Research Letters}, 38(1):39--46.

\bibitem[Filippi et~al., 2023]{filippi2023incorporating}
Filippi, C., Guastaroba, G., Peirano, L., and Speranza, M.~G. (2023).
\newblock {Incorporating time-dependent demand patterns in the optimal location
  of capacitated charging stations}.
\newblock {\em Transportation Research Part C: Emerging Technologies},
  152:104145.

\bibitem[Fischetti et~al., 2016]{fischetti2016benders}
Fischetti, M., Ljubi{\'c}, I., and Sinnl, M. (2016).
\newblock {Benders decomposition without separability: A computational study
  for capacitated facility location problems}.
\newblock {\em European Journal of Operational Research}, 253(3):557--569.

\bibitem[Fischetti et~al., 2017]{fischetti2017redesigning}
Fischetti, M., Ljubi{\'c}, I., and Sinnl, M. (2017).
\newblock {Redesigning Benders decomposition for large-scale facility
  location}.
\newblock {\em Management Science}, 63(7):2146--2162.

\bibitem[Fisk, 1980]{fisk80}
Fisk, C. (1980).
\newblock {Some developments in equilibrium traffic assignment}.
\newblock {\em Transportation Research Part B: Methodological}, 14(3):243--255.

\bibitem[{Government of British Columbia}, 2025]{cleanbcPublicCharger2025}
{Government of British Columbia} (2025).
\newblock {CleanBC} go electric public charger program.
\newblock Link:
  \url{https://www2.gov.bc.ca/gov/content/industry/electricity-alternative-energy/transportation-energies/clean-transportation-policies-programs/clean-energy-vehicle-program/dcfc-program/hydrogen-fuelling-52518}.
  Accessed: 2025-05-26.

\bibitem[{Government of Québec}, 2023]{quebecEVstrategy2023}
{Government of Québec} (2023).
\newblock Québec's electric vehicle charging strategy 2023–2030.
\newblock Link:
  \url{https://www.quebec.ca/en/government/policies-orientations/quebec-electric-vehicle-charging-strategy}.
  Accessed: 2025-05-26.

\bibitem[Guillet and Schiffer, 2025]{guillet2025coordinating}
Guillet, M. and Schiffer, M. (2025).
\newblock {Coordinating charging request allocation between self-interested
  navigation service platforms}.
\newblock {\em INFORMS Journal on Computing}, 37(2):293--314.

\bibitem[Haase and M{\"u}ller, 2013]{haase2013management}
Haase, K. and M{\"u}ller, S. (2013).
\newblock {Management of school locations allowing for free school choice}.
\newblock {\em Omega}, 41(5):847--855.

\bibitem[{International Energy Agency}, 2024]{iea_global_ev_outlook_2024}
{International Energy Agency} (2024).
\newblock Global {EV} outlook 2024.
\newblock First published in February 2024.

\bibitem[{Jalili Marand} and Hoseinpour, 2024]{JALILIMARAND2024442}
{Jalili Marand}, A. and Hoseinpour, P. (2024).
\newblock {A congested facility location problem with strategic customers}.
\newblock {\em European Journal of Operational Research}, 318(2):442--456.

\bibitem[K\i{}nay et~al., 2023]{Kinay2023ChargingSL}
K\i{}nay, O.~B., Gzara, F., and Alumur, S.~A. (2023).
\newblock Charging station location and sizing for electric vehicles under
  congestion.
\newblock {\em Transportation Science}, 57(6):1433–1451.

\bibitem[Kullback and Leibler, 1951]{kullback1951information}
Kullback, S. and Leibler, R.~A. (1951).
\newblock {On information and sufficiency}.
\newblock {\em The annals of mathematical statistics}, 22(1):79--86.

\bibitem[Lamontagne et~al., 2024]{lamontagne2024accelerated}
Lamontagne, S., Carvalho, M., and Atallah, R. (2024).
\newblock {Accelerated Benders decomposition and local branching for dynamic
  maximum covering location problems}.
\newblock {\em Computers \& Operations Research}, 167:106673.

\bibitem[Lamontagne et~al., 2023]{Lamontagne2022OptimisingEV}
Lamontagne, S., Carvalho, M., Frejinger, E., Gendron, B., Anjos, M.~F., and
  Atallah, R. (2023).
\newblock Optimising electric vehicle charging station placement using advanced
  discrete choice models.
\newblock {\em INFORMS Journal on Computing}, 35(5):1195–1213.

\bibitem[Laporte et~al., 2019]{Laporte2019}
Laporte, G., Nickel, S., and Saldanha-da Gama, F. (2019).
\newblock {\em Introduction to Location Science}, pages 1--21.
\newblock Springer International Publishing, Cham.

\bibitem[Lin et~al., 2023]{LIN2023106175}
Lin, Y.~H., Tian, Q., and Liu, S. (2023).
\newblock {Service expansion for chained business facilities under congestion
  and market competition}.
\newblock {\em Computers \& Operations Research}, 153:106175.

\bibitem[Liu et~al., 2023]{Liu01032023}
Liu, B., Pantelidis, T.~P., Tam, S., and and, J. Y. J.~C. (2023).
\newblock {An electric vehicle charging station access equilibrium model with
  M/D/C queueing}.
\newblock {\em International Journal of Sustainable Transportation},
  17(3):228--244.

\bibitem[Ma et~al., 2020]{Ma2020-iq}
Ma, H., Guan, X., and Wang, L. (2020).
\newblock A single-facility competitive location problem in the plane based on
  customer choice rules.
\newblock {\em Journal of Data, Information and Management}, 2(4):323--336.

\bibitem[Marianov et~al., 2008]{Marianov2008FacilityLF}
Marianov, V., Ríos, M., and Icaza, M.~J. (2008).
\newblock {Facility location for market capture when users rank facilities by
  shorter travel and waiting times}.
\newblock {\em European Journal of Operational Research}, 191(1):32--44.

\bibitem[McFadden, 1974]{mcfadden_conditional_1974}
McFadden, D. (1974).
\newblock {Conditional logit analysis of qualitative choice behavior}.
\newblock In Zarembka, P., editor, {\em Fontiers in {Econometrics}}, pages
  105--142. Academic press, New York.

\bibitem[Parent et~al., 2024]{Parent2023MaximumFF}
Parent, P., Carvalho, M., Anjos, M., and Atallah, R. (2024).
\newblock {Maximum flow‐based formulation for the optimal location of
  electric vehicle charging stations}.
\newblock {\em Networks}, 84.

\bibitem[Plastria, 2001]{PLASTRIA2001461}
Plastria, F. (2001).
\newblock {Static competitive facility location: An overview of optimisation
  approaches}.
\newblock {\em European Journal of Operational Research}, 129(3):461--470.

\bibitem[Smith, 2008]{Smith2008MGCkPM}
Smith, J.~M. (2008).
\newblock {M/G/c/K performance models in manufacturing and service systems}.
\newblock {\em Asia-Pacific Journal of Operational Research}, 25(04):531--561.

\bibitem[Stewart, 2009]{stewart09}
Stewart, W.~J. (2009).
\newblock {\em {Probability, Markov Chains, Queues, and Simulation: The
  Mathematical Basis of Performance Modeling}}.
\newblock Princeton University Press.

\bibitem[Sugishita et~al., 2025]{sugishita2025fair}
Sugishita, N., Sevim, I., Carvalho, M., Dems, A., and Atallah, R. (2025).
\newblock Fair network design problem: An application to {EV} charging station
  capacity expansion.
\newblock \url{https://www.optimization-online.org/DB_HTML/2025/01/xxxx.html}.
\newblock Optimization Online preprint.

\bibitem[Ulloa et~al., 2024]{ulloa2024logistics}
Ulloa, D.~P., Frejinger, E., and Gendron, B. (2024).
\newblock {A logistics provider’s profit maximization facility location
  problem with random utility maximizing followers}.
\newblock {\em Computers \& Operations Research}, 167:106649.

\bibitem[Xiao et~al., 2020]{XIAO2020101317}
Xiao, D., An, S., Cai, H., Wang, J., and Cai, H. (2020).
\newblock {An optimization model for electric vehicle charging infrastructure
  planning considering queuing behavior with finite queue length}.
\newblock {\em Journal of Energy Storage}, 29:101317.

\bibitem[Xie et~al., 2018]{xie2018long}
Xie, F., Liu, C., Li, S., Lin, Z., and Huang, Y. (2018).
\newblock {Long-term strategic planning of inter-city fast charging
  infrastructure for battery electric vehicles}.
\newblock {\em Transportation Research Part E: Logistics and Transportation
  Review}, 109:261--276.

\bibitem[Xie and Xie, 2016]{XIE2016406}
Xie, J. and Xie, C. (2016).
\newblock {New insights and improvements of using paired alternative segments
  for traffic assignment}.
\newblock {\em Transportation Research Part B: Methodological}, 93:406--424.

\bibitem[Yang, 2018]{YANG2018189}
Yang, W. (2018).
\newblock {A user-choice model for locating congested fast charging stations}.
\newblock {\em Transportation Research Part E: Logistics and Transportation
  Review}, 110:189--213.

\bibitem[Zhang et~al., 2023]{Zhang04032023}
Zhang, B., Zhao, M., and and, X.~H. (2023).
\newblock {Location planning of electric vehicle charging station with users’
  preferences and waiting time: multi-objective bi-level programming model and
  HNSGA-II algorithm}.
\newblock {\em International Journal of Production Research}, 61(5):1394--1423.

\end{thebibliography}

\newpage
\appendix
\section{Detailed results for Section~\ref{subsec:queues_comparison}}
\label{appendix:queues_comparison}

We presented supplementary results to those of Section~\ref{subsec:queues_comparison}. As in that section, we consider queues with total service rate of 40.

\begin{figure}[H]
  \centering

  \includegraphics[width=\textwidth]{queues_comparison_B_0_s_1_2.png}
  \vspace{0.5cm}
  \caption*{(a) Buffer = 0}

  \includegraphics[width=\textwidth]{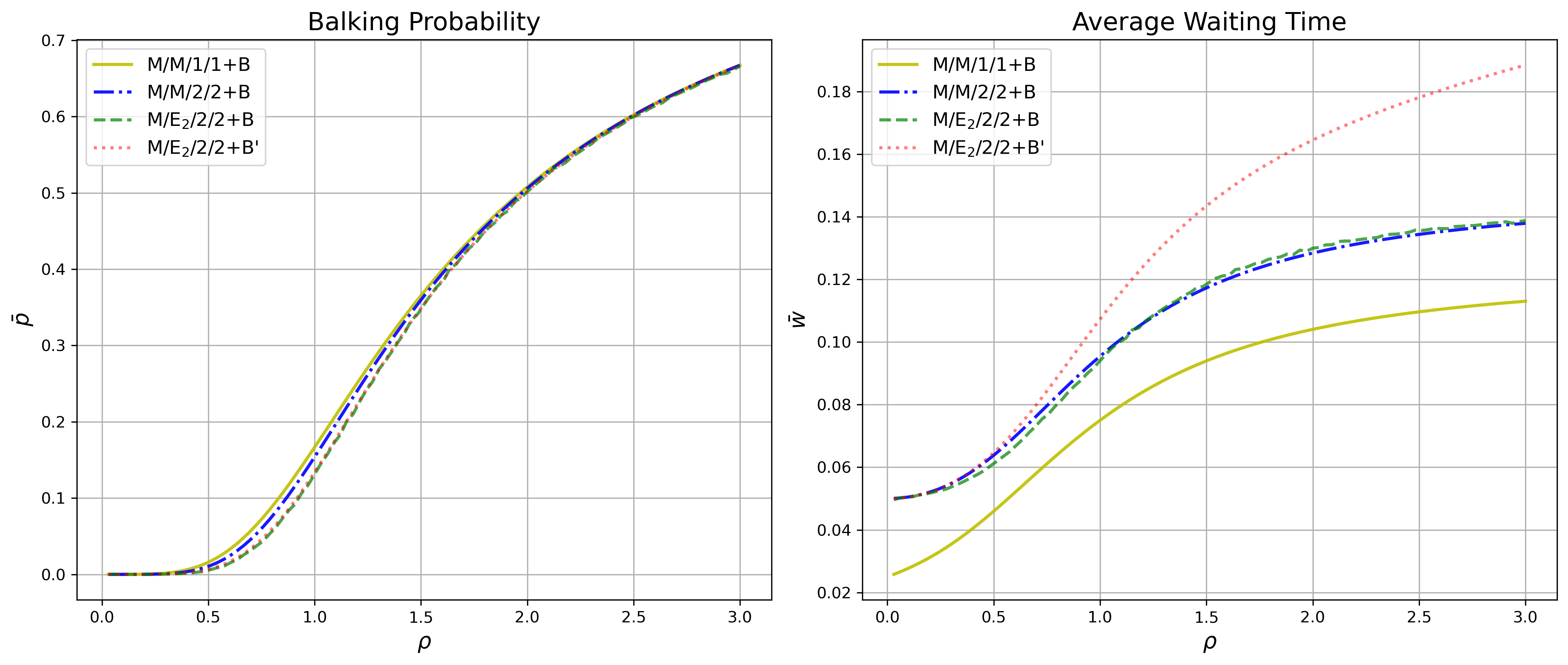}
  \caption*{(b) Buffer = 4}

  \caption{Comparison of queueing performance between single- and multi-server queues with $s = 2$ (part 1).}
  \label{fig:queue_comparison_part1}
\end{figure}

\begin{figure}[H]
  \centering

  \includegraphics[width=\textwidth]{queues_comparison_B_10_s_1_2.png}
  \vspace{0.5cm}
  \caption*{(c) Buffer = 10}

  \includegraphics[width=\textwidth]{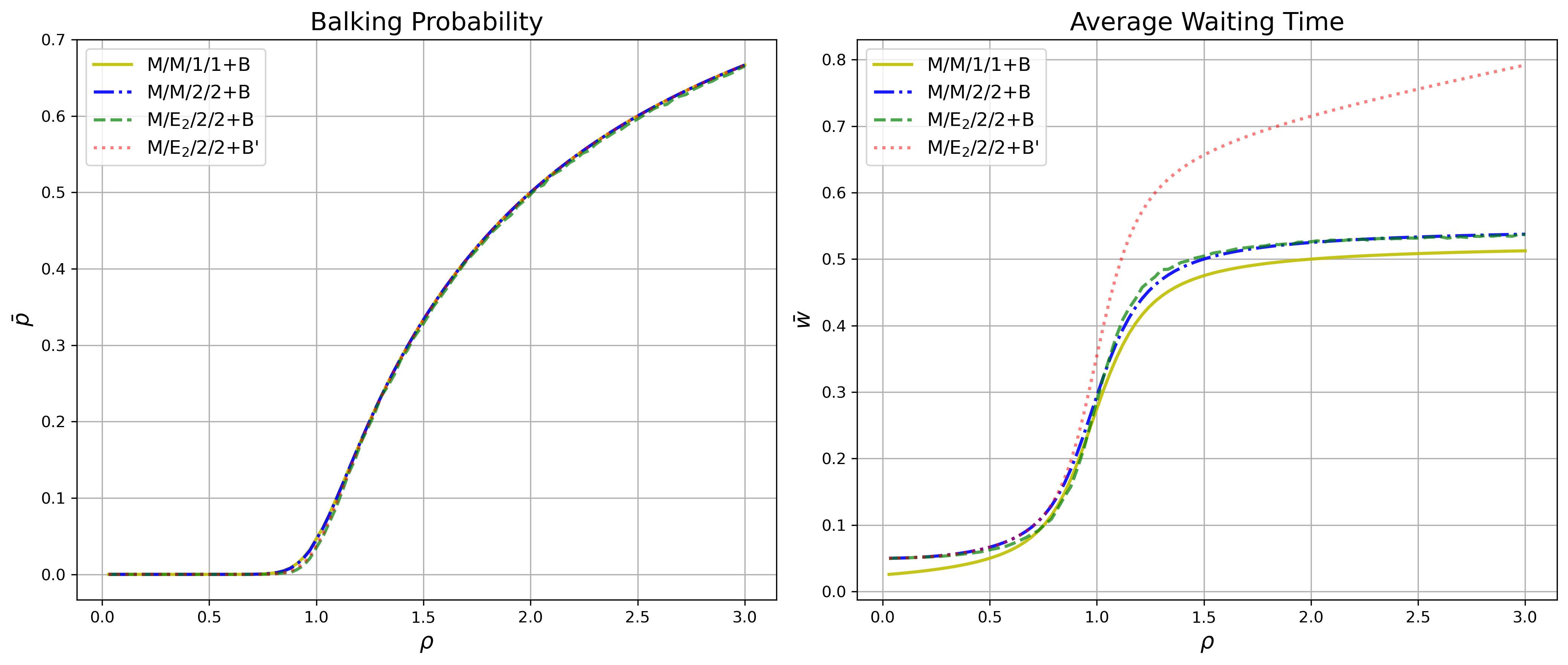}
  \caption*{(d) Buffer = 20}

  \caption{Comparison of queueing performance between single- and multi-server queues with $s = 2$ (part 2).}
  \label{fig:queue_comparison_part2}
\end{figure}

\begin{figure}[H]
  \centering

  \includegraphics[width=\textwidth]{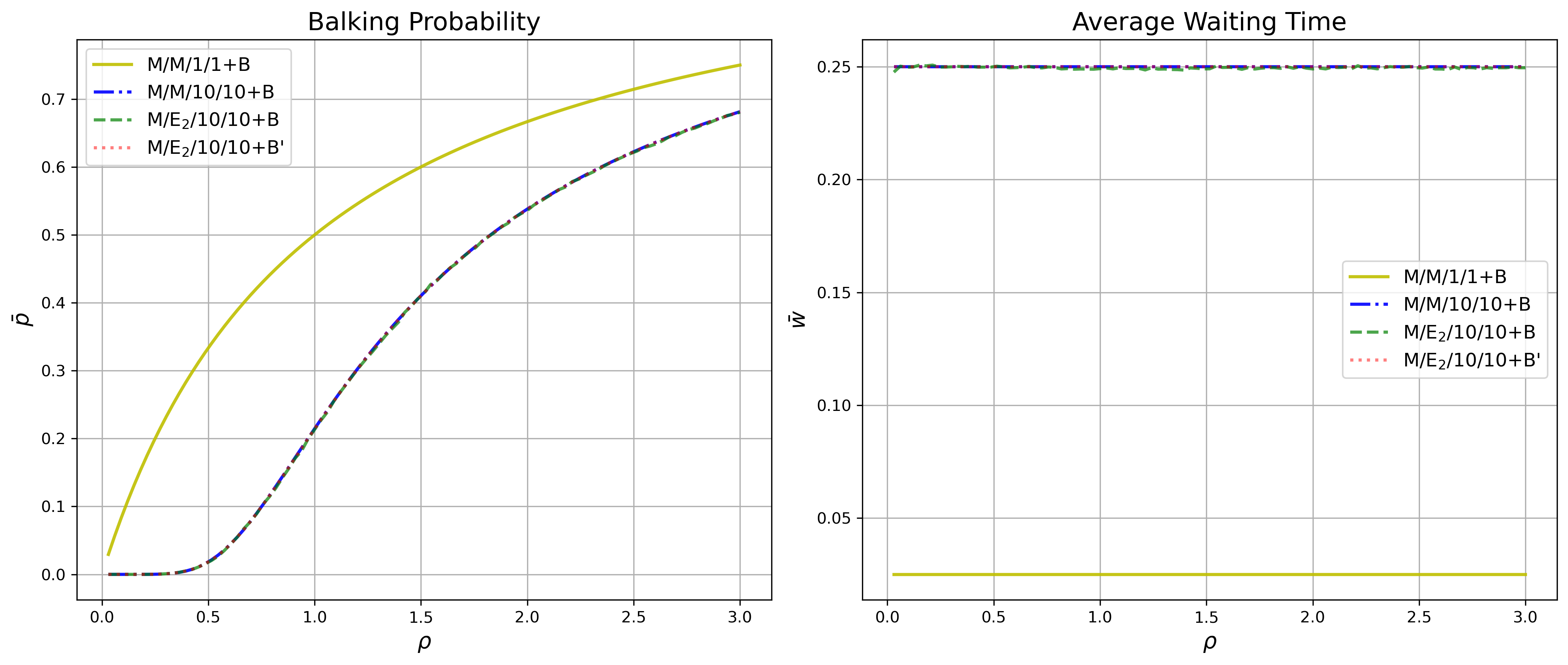}
  \vspace{0.5cm}
  \caption*{(a) Buffer = 0}

  \includegraphics[width=\textwidth]{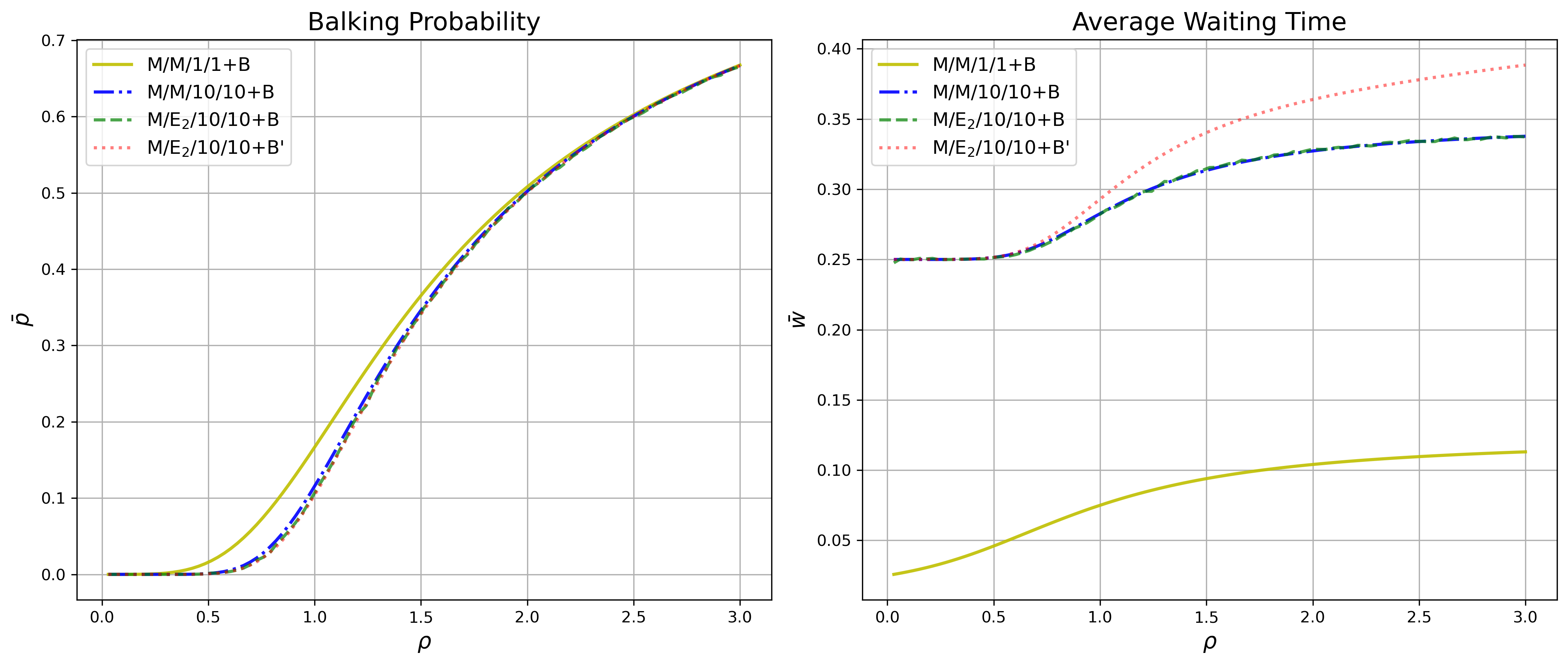}
  \caption*{(b) Buffer = 4}

  \caption{Comparison of queueing performance between single- and multi-server queues with $s = 10$ (part 1).}
  \label{fig:queue_comparison_s10_part1}
\end{figure}

\begin{figure}[H]
  \centering

  \includegraphics[width=\textwidth]{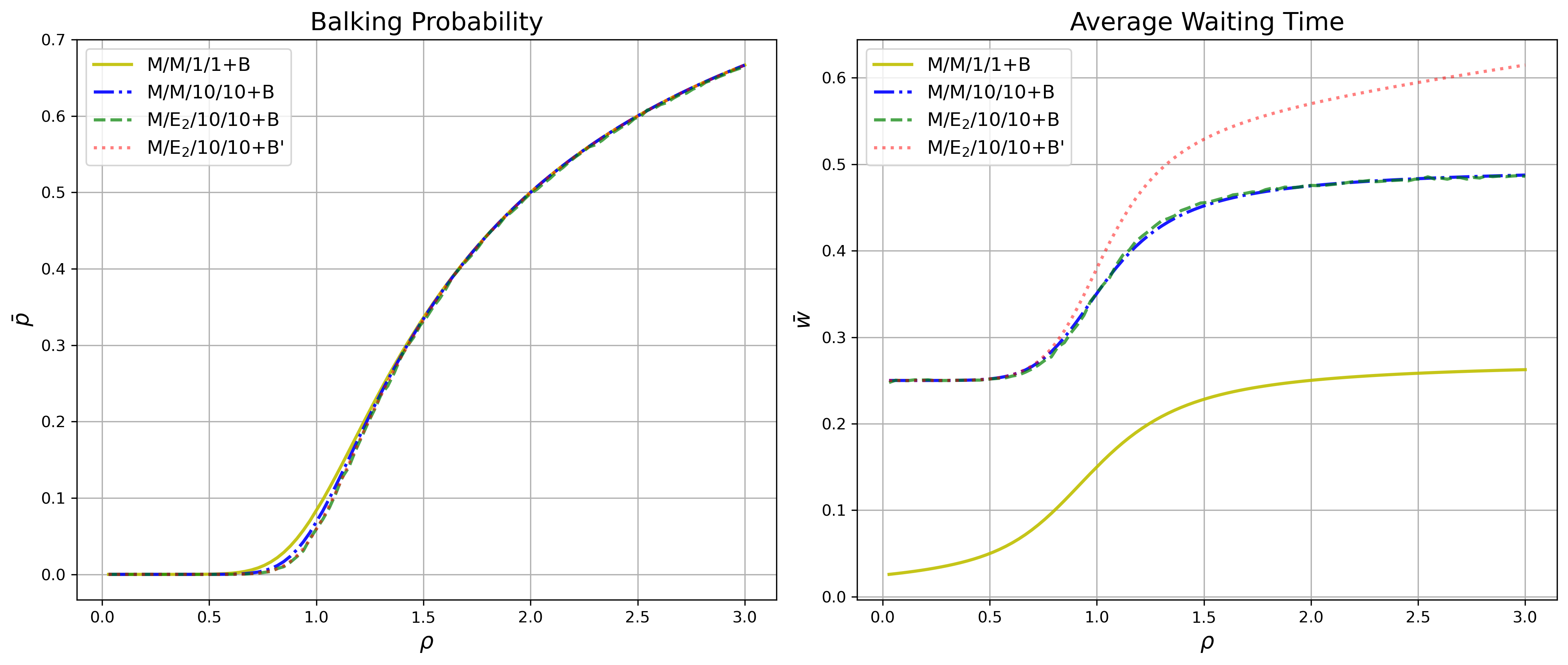}
  \vspace{0.5cm}
  \caption*{(c) Buffer = 10}

  \includegraphics[width=\textwidth]{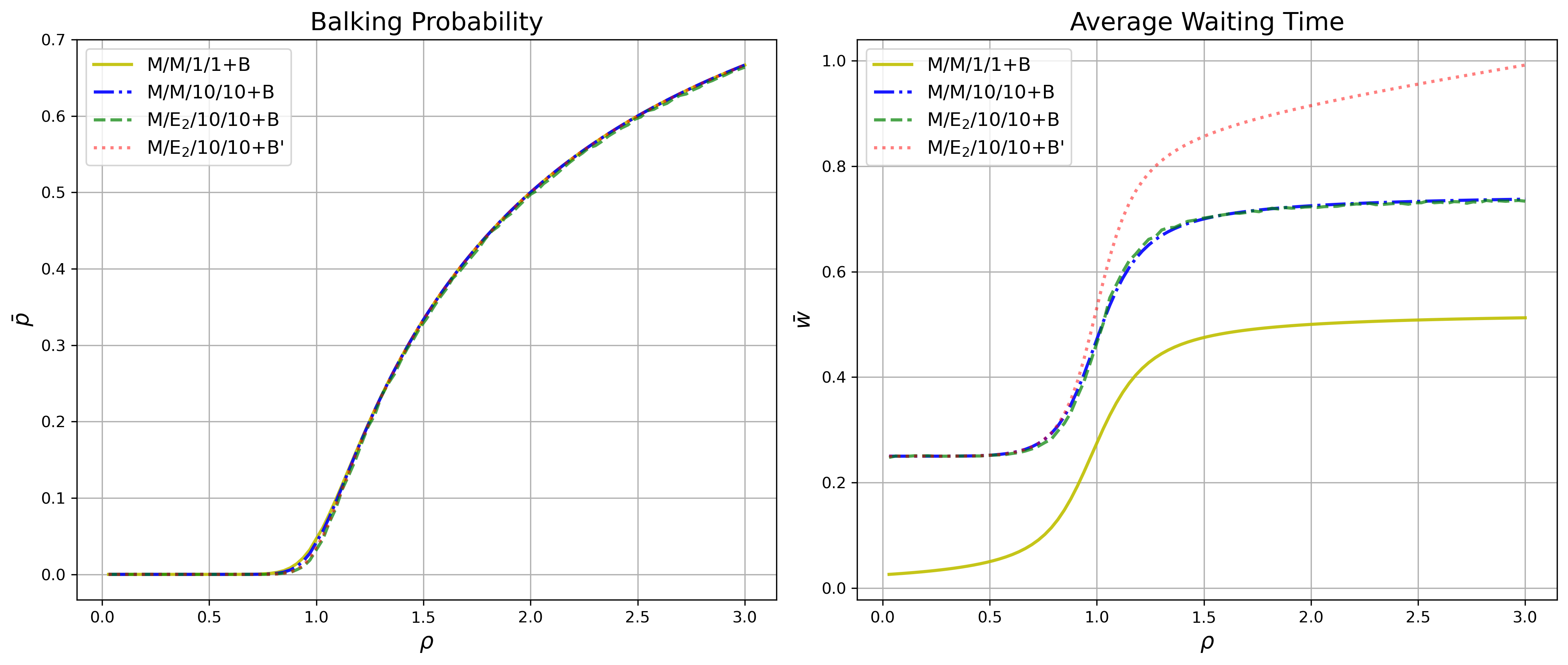}
  \caption*{(d) Buffer = 20}

  \caption{Comparison of queueing performance between single- and multi-server queues with $s = 10$ (part 2).}
  \label{fig:queue_comparison_s10_part2}
\end{figure}

The performance metrics—balking probability and average waiting time—are shown in Figures~\ref{fig:queue_comparison_part1} and~\ref{fig:queue_comparison_part2} for queue models with two servers ($s=2$), and in Figures~\ref{fig:queue_comparison_s10_part1} and~\ref{fig:queue_comparison_s10_part2} for models with ten servers ($s=10$).

\section{Proofs supporting Proposition~\ref{prop:llp_convex}}
\label{appendix:convexity_proof}

To prove Proposition~\ref{prop:llp_convex}, we first establish Lemmata~\ref{lem:mmsk_p}--\ref{lem:mmsk_w}, whose proofs are provided below.

\begin{lemma}\label{lem:mmsk_p}
Let $s$ and $K$ be positive integers such that $s \le K$ and let $\ServiceRate$ be a positive real number.
The probability of balking $\bar{p}$ of the $M/M/s/K$ queue with service rate $\ServiceRate$ is increasing in $\lambda$.
\end{lemma}

\begin{proof}
Let $\rho = \ArrivalRate / (\NServers \ServiceRate)$. From equation~\eqref{eq:mmsk_probability_of_n_users}, we obtain an alternative expression for $\bar{p}$:
\begin{align*}
    \bar{p} = \dfrac{s^{s}}{s!}\left[\sum_{m=0}^{s-1} \dfrac{s^m \rho^{m-K}}{m!} + \dfrac{s^{s}}{s!}\sum_{m=s}^K \rho^{m-K}\right]^{-1}.
\end{align*}

The derivative of $\bar{p}$ with respect to $\lambda$ is:
\begin{align*}
    \pdv{\bar{p}}{\lambda} = \pdv{\bar{p}}{\rho} \pdv{\rho}{\lambda} = \pdv{\bar{p}}{\rho} \dfrac{1}{s \mu}.
\end{align*}

Also,
\begin{align*}
    \pdv{\bar{p}}{\rho} = - \dfrac{s^{s}}{s!} \dfrac{\left[ \sum_{m=0}^{s-1} (m-K) \dfrac{s^m \rho^{m-K-1}}{m!} + \dfrac{s^{s}}{s!}\sum_{m=s}^{K-1} (m-K) \rho^{m-K-1} \right]}{\left[ \sum_{m=0}^{s-1} \dfrac{s^m \rho^{m-K}}{m!} + \dfrac{s^{s}}{s!}\sum_{m=s}^K \rho^{m-K} \right]^{2}} \geq 0.
\end{align*}

It follows that $\pdv{\bar{p}}{\lambda} \geq 0$, and therefore, $\bar{p}$ is increasing in $\lambda$.
\end{proof}

Next, we provide a technical lemma that will be used to prove the monotonicity of the average waiting time $\bar{w}$.

\begin{lemma}
\label{lemma:derivative_of_waiting_time_is_positive}
Let $s$ and $K$ be positive integers such that $s \le K$.
Then, we have $E_{s K}(\tau) \ge 0$ for any $\tau \in \mathbb{R}_+$, where
\begin{align*}
E_{s K}(\tau)
& := \left(
\sum_{n = 2}^{s - 1} \frac{\tau^{n - 2}}{(n - 2)!}
+ \sum_{n = s}^K n(n-1)\frac{\tau^{n - 2}}{s! s^{n - s}}
\right)
\left(
\sum_{n = 0}^{s - 1} \frac{\tau^n}{n!}
+ \sum_{n = s}^{K - 1} \frac{\tau^n}{s! s^{n - s}}
\right)
\\
& \qquad -
\left(
\sum_{n = 1}^{s - 1} \frac{\tau^{n - 1}}{(n - 1)!}
+ \sum_{n = s}^K n \frac{\tau^{n - 1}}{s! s^{n - s}}
\right)
\left(
\sum_{n = 1}^{s - 1} \frac{\tau^{n - 1}}{(n - 1)!}
+ \sum_{n = s}^{K - 1} n \frac{\tau^{n - 1}}{s! s^{n - s}}
\right).
\end{align*}
\end{lemma}

\begin{proof}
Let $s$ be any positive integer.
We prove the following holds for any $\InductionMainK$ using induction:
$$
E_{s \InductionMainK}(\tau) \ge 0, \qquad \forall \tau \ge 0.
$$

\textbf{Base case:}
The base case ($\InductionMainK = s$) follows since for any $\tau \ge 0$
\begin{align*}
E_{s s}(\tau)
& = \left(
\sum_{n = 0}^{s - 3} \frac{\tau^n}{n!}
+ s(s-1)\frac{\tau^{s - 2}}{s!}
\right)
\left(
\sum_{n = 0}^{s - 1} \frac{\tau^n}{n!}
\right)
-
\left(
\sum_{n = 0}^{s - 2} \frac{\tau^n}{n!}
+ s \frac{\tau^{s - 1}}{s!}
\right)
\left(
\sum_{n = 0}^{s - 2} \frac{\tau^n}{n!}
\right) \\
& = \left(
\sum_{n = 0}^{s - 2} \frac{\tau^n}{n!}
\right)
\left(
\sum_{n = 0}^{s - 1} \frac{\tau^n}{n!}
\right)
-
\left(
\sum_{n = 0}^{s - 1} \frac{\tau^n}{n!}
\right)
\left(
\sum_{n = 0}^{s - 2} \frac{\tau^n}{n!}
\right)
= 0.
\end{align*}

\textbf{Inductive case:}

Assuming $E_{s \, \InductionDummyK - 1}(\tau) \ge 0$, $\forall \tau \ge 0$, for some $\InductionDummyK$, we show $E_{s \InductionDummyK}(\tau) \ge 0$, $\forall \tau \ge 0$.

Direct calculation gives
\begin{align*}
& \left(\frac{\tau^{\InductionDummyK - 2}}{s! s^{\InductionDummyK - s}}\right)^{-1} (E_{s \InductionDummyK}(\tau) - E_{s \, \InductionDummyK - 1}(\tau)) \\ 
& = 
\InductionDummyK(\InductionDummyK-1)
\left(
\sum_{n = 0}^{s - 1} \frac{\tau^n}{n!}
+ \sum_{n = s}^{\InductionDummyK - 2} \frac{\tau^n}{s! s^{n - s}}
\right)
 +
\tau s \left(
\sum_{n = 0}^{s - 3} \frac{\tau^n}{n!}
+ \sum_{n = s}^{\InductionDummyK - 1} n(n-1)\frac{\tau^{n - 2}}{s! s^{n - s}}
\right)
\\
& \qquad -
\InductionDummyK \tau
\left(
\sum_{n = 0}^{s - 2} \frac{\tau^n}{n!}
+ \sum_{n = s}^{\InductionDummyK - 2} n \frac{\tau^{n - 1}}{s! s^{n - s}}
\right)
-
(\InductionDummyK - 1) s 
\left(
\sum_{n = 0}^{s - 2} \frac{\tau^n}{n!}
+ \sum_{n = s}^{\InductionDummyK - 1} n \frac{\tau^{n - 1}}{s! s^{n - s}} \right) \\
& =: h_{s \InductionDummyK}(\tau).
\end{align*}
Thus, we will show $h_{s \InductionDummyK}(\tau) \ge 0$ for all $\tau \ge 0$, which implies $E_{s \InductionDummyK}(\tau) \ge 0$ for all $\tau \ge 0$.
Distribute the product to get
\begin{align*}
h_{s \InductionDummyK}(\tau)
& =
\InductionDummyK(\InductionDummyK-1)
\sum_{n = 0}^{s - 1} \frac{\tau^n}{n!}
+
\InductionDummyK(\InductionDummyK-1) \sum_{n = s}^{\InductionDummyK - 2} \frac{\tau^n}{s! s^{n - s}}
+
\tau s
\sum_{n = 0}^{s - 3} \frac{\tau^n}{n!}
+
\tau s
\sum_{n = s}^{\InductionDummyK - 1} n(n-1)\frac{\tau^{n - 2}}{s! s^{n - s}}
\\
& \qquad -
\InductionDummyK \tau
\sum_{n = 0}^{s - 2} \frac{\tau^n}{n!}
- 
\InductionDummyK \tau
\sum_{n = s}^{\InductionDummyK - 2} n \frac{\tau^{n - 1}}{s! s^{n - s}}
-
(\InductionDummyK - 1) s 
\sum_{n = 0}^{s - 2} \frac{\tau^n}{n!}
-
(\InductionDummyK - 1) s 
\sum_{n = s}^{\InductionDummyK - 1} n \frac{\tau^{n - 1}}{s! s^{n - s}}.
\end{align*}
Rearrange the terms in the summations to get
\begin{align*}
h_{s \InductionDummyK}(\tau)
& =
\InductionDummyK(\InductionDummyK-1)\left(
\sum_{n = 0}^{s - 3} \frac{\tau^n}{n!}
+
\frac{\tau^{s - 2}}{(s - 2)!}
+
\frac{\tau^{s - 1}}{(s - 1)!}
\right)
\\
& \qquad
+
\InductionDummyK(\InductionDummyK-1)
\sum_{n = s}^{\InductionDummyK - 2} \frac{\tau^n}{s! s^{n - s}}
+
 \tau s
\sum_{n = 0}^{s - 3} \frac{\tau^n}{n!}
+
\tau s
\left(
\sum_{n = s + 1}^{\InductionDummyK - 1} n(n-1)\frac{\tau^{n - 2}}{s! s^{n - s}}
+
s (s - 1) \frac{\tau^{s - 2}}{s!}
\right)
\\
& \qquad -
\left(
 \InductionDummyK \tau
\sum_{n = 0}^{s - 3} \frac{\tau^n}{n!}
+
 \InductionDummyK \tau
\frac{\tau^{s - 2}}{(s - 2)!}
\right)
-
 \InductionDummyK \tau
\sum_{n = s}^{\InductionDummyK - 2} n \frac{\tau^{n - 1}}{s! s^{n - s}}
\\
& \qquad -
 (\InductionDummyK - 1) s 
 \left(
\sum_{n = 0}^{s - 3} \frac{\tau^n}{n!}
-
\frac{\tau^{s - 2}}{(s - 2)!}
\right)
-
 (\InductionDummyK - 1) s 
 \left(
\sum_{n = s + 1}^{\InductionDummyK - 1} n \frac{\tau^{n - 1}}{s! s^{n - s}}
-
s
\frac{\tau^{s - 1}}{s!}
\right)
\\
& =
 (\InductionDummyK - s) (\InductionDummyK - \tau - 1)
\sum_{n = 0}^{s - 3} \frac{\tau^n}{n!}
+
 (\InductionDummyK - s) (\InductionDummyK - 1) 
\frac{\tau^{s - 2}}{(s - 2)!} 
+
 (\InductionDummyK - s)^2
\frac{\tau^{s - 1}}{(s - 1)!}
\\
& \qquad +
 \sum_{n = s}^{\InductionDummyK - 2} (\InductionDummyK - n - 1)^2 \frac{\tau^n}{s! s^{n - s}}
 \\
 &\ge
 (\InductionDummyK - s) (\InductionDummyK - \tau - 1)
\sum_{n = 0}^{s - 3} \frac{\tau^n}{n!}
+
 (\InductionDummyK - s) (\InductionDummyK - 1) 
\frac{\tau^{s - 2}}{(s - 2)!} 
:= h'_{s \InductionDummyK}(\tau).
\end{align*}
Now observe that
\begin{align*}
h'_{s \InductionDummyK}(\tau)
& =
\Big(
(\InductionDummyK - s) (\InductionDummyK - 1)
- (\InductionDummyK - s) \tau
\Big)
\sum_{n = 0}^{s - 3} \frac{\tau^n}{n!}
+
(\InductionDummyK - s) (\InductionDummyK - 1) 
\frac{\tau^{s - 2}}{(s - 2)!} 
\\
& =
(\InductionDummyK - s) (\InductionDummyK - 1)
\left(
1
+
\sum_{n = 1}^{s - 3} \frac{\tau^n}{n!} 
\right)
-
(\InductionDummyK - s)
\left(
\sum_{n = 0}^{s - 4} \frac{\tau^{n + 1}}{n!}
+
\frac{\tau^{s - 2}}{(s - 3)!}
\right)
\\
& \qquad 
+
(\InductionDummyK - s) (\InductionDummyK - 1) 
\frac{\tau^{s - 2}}{(s - 2)!} 
\\
& =
(\InductionDummyK - s) (\InductionDummyK - 1)
\left(
1
+
\sum_{n = 0}^{s - 4} \frac{\tau^{n + 1}}{(n + 1)!} 
\right)
-
(\InductionDummyK - s)
\left(
\sum_{n = 0}^{s - 4} \frac{\tau^{n + 1}}{n!}
+
\frac{\tau^{s - 2}}{(s - 3)!}
\right)
\\
& \qquad 
+
(\InductionDummyK - s) (\InductionDummyK - 1) 
\frac{\tau^{s - 2}}{(s - 2)!} 
\\
& \ge
(\InductionDummyK - s)
\sum_{n = 0}^{s - 4} (\InductionDummyK - n - 2) \frac{\tau^{n + 1}}{(n + 1)!}
+
(\InductionDummyK - s) (\InductionDummyK - s + 1) 
\frac{\tau^{s - 2}}{(s - 2)!} 
\ge 0.
\end{align*}
Therefore, we have $h_{s \InductionDummyK}(\tau) \ge 0$ for all $\tau \ge 0$, implying $E_{s \InductionDummyK}(\tau) \ge 0$ for all $\tau \ge 0$.

\textbf{Conclusion:} Since both the base case and the induction step have been proven as true, by mathematical induction $E_{s \InductionMainK}(\tau) \ge 0$ for all $\tau \ge0, \InductionMainK \ge s$.
\end{proof}

\begin{lemma}\label{lem:mmsk_w}
Let $s$ and $K$ be positive integers such that $s \le K$ and let $\ServiceRate$ be a positive real number.
The average waiting time $\bar{w}$ of the $M/M/s/K$ queue with service rate $\ServiceRate$ is increasing in $\lambda$.
\end{lemma}

\begin{proof}Equation~\eqref{eq:queueing_average_waiting_time} gives
\begin{align*}
    \bar{w} &= \dfrac{\sum_{n=0}^K np^{(n)}}{\lambda(1-\bar{p})}
    =  \dfrac{\sum_{n=1}^{s-1} \dfrac{\tau^{n-1}}{(n-1)!} + \sum_{n=s}^{K} n  \dfrac{\tau^{n-1}}{s! s^{n-s}}}{\mu \left(\sum_{m=0}^{s-1}  \dfrac{\tau^m}{m!} + \sum_{m=s}^{K-1} \dfrac{\tau^m}{s! s^{m-s}} \right)},
\end{align*}
where $\tau = \lambda / \mu$.
Thus,
\begin{align*}
    \pdv{\bar{w}}{\lambda} &= 
    \pdv{\bar{w}}{\tau}\pdv{\tau}{\lambda}
    \\
    &=
    \dfrac{1}{\mu^2}
    \left\{
   \left(\sum_{m=0}^{s-1}  \dfrac{\tau^m}{m!} + \sum_{m=s}^{K-1} \dfrac{\tau^m}{s! s^{m-s}} \right)
    \right\}^{-4} \\
    & \qquad
    \Biggl\{
    \left(\sum_{n=2}^{s-1} \dfrac{\tau^{n-2}}{(n-2)!} + \sum_{n=s}^{K} n(n-1)  \dfrac{\tau^{n-2}}{s! s^{n-s}} \right) 
    \left(\sum_{m=0}^{s-1}  \dfrac{\tau^m}{m!} + \sum_{m=s}^{K-1} \dfrac{\tau^m}{s! s^{m-s}} \right) \\
    & \qquad\qquad
    - \left( \sum_{n=1}^{s-1} \dfrac{\tau^{n-1}}{(n-1)!} + \sum_{n=s}^{K} n  \dfrac{\tau^{n-1}}{s! s^{n-s}} \right)
    \left( \sum_{m=1}^{s-1}  \dfrac{\tau^{m-1}}{(m-1)!} + \sum_{m=s}^{K-1} m \dfrac{\tau^{m-1}}{s! s^{m-s}} \right)
    \Biggr\}.
\end{align*}
Using Lemma~\ref{lemma:derivative_of_waiting_time_is_positive} and $\tau = \lambda / \mu$, we conclude that this is nonnegative for any $\lambda \ge 0$, which implies the desired result.

\end{proof}

\noindent\textbf{Proposition~\ref{prop:llp_convex}.}
For the $M/M/s/K$ queue, problem~\eqref{eq:llp} is convex.

\begin{proof}
From Lemmata~\ref{lem:mmsk_p} and~\ref{lem:mmsk_w}, it follows that functions~\eqref{func:f^p_LLP} and~\eqref{func:f^w_LLP} are convex in $\ArrivalRate$. Together with the convexity of function~\eqref{func:f^l_LLP}, this implies that the objective function of problem~\eqref{eq:llp} is convex in the variables $\DisaggregatedArrivalRate$ and $\ArrivalRate$. Therefore, problem~\eqref{eq:llp} is convex.
\end{proof}
\end{document}